\documentclass[aps,pra,twocolumn,showpacs,superscriptaddress,amsmath,amssymb]{revtex4-1}

\usepackage{graphicx}
\usepackage{hyperref}
\usepackage{color}
\usepackage{mathdots}
\usepackage{microtype}

\newcommand{\fig}[1]{Fig.~\ref{fig:#1}}

\newcommand{\REFEREE}[1]{\textcolor{black}{#1}}
\renewcommand{\Re}{\mathrm{Re}}
\renewcommand{\Im}{\mathrm{Im}}
\newcommand{\defined}{=}
\renewcommand{\ell}{m}

\hypersetup{bookmarksnumbered=true,
  bookmarksopen=true,
  citecolor=blue,
  colorlinks=true,
  hypertexnames=true,
  linkcolor=blue,
  linktocpage=true,
  pdfauthor={Sofi Esterhazy, David Liu, Matthias Liertzer, Jens Markus Melenk, Alexander Cerjan, Li Ge, A. Douglas Stone, Steven G. Johnson, Stefan Rotter},
  pdfhighlight=/I,
  pdfpagemode=UseNone,
  pdftitle={Scalable numerical approach for the steady-state ab initio laser theory},
  urlcolor=blue}

\newcommand\BPsi{\mathbf{\Psi}}
\newcommand\bark{\bar{k}}
\newcommand\barpsi{\bar{\mathbf{\Psi}}}

\begin{document}

\title{Scalable numerical approach for the steady-state \emph{ab initio} laser theory}

\author{S.~Esterhazy}
\affiliation{\textls[-20]{Institute for Analysis and Scientific Computing, Vienna University of
              Technology, A-1040 Vienna, Austria, EU}}

\author{D.~Liu}
\affiliation{ Department of Physics, Massachusetts Institute of Technology,
              Cambridge, Massachusetts 02139, USA}

\author{M.~Liertzer}
\affiliation{ Institute for Theoretical Physics, Vienna University of
              Technology, A-1040 Vienna, Austria, EU}

\author{A.~Cerjan}
\affiliation{ Department of Applied Physics, Yale University, New Haven, Connecticut 06520, USA}

\author{L.~Ge}
\affiliation{\textls[-20]{Department of Engineering Science and Physics, College of Staten Island, CUNY, New York, New York 10314, USA}}
\affiliation{ The Graduate Center, CUNY, New York, New York 10016, USA }

\author{K.G.~Makris}
\affiliation{ Institute for Theoretical Physics, Vienna University of
              Technology, A-1040 Vienna, Austria, EU}
\affiliation{\textls[-20]{Department of Electrical Engineering, Princeton University, Princeton, New Jersey 08544, USA}}

\author{A.D.~Stone}
\affiliation{ Department of Applied Physics, Yale University, New Haven, Connecticut 06520, USA}

\author{J.M.~Melenk}
\affiliation{\textls[-20]{Institute for Analysis and Scientific Computing, Vienna University of
              Technology, A-1040 Vienna, Austria, EU}}

\author{S.G.~Johnson}
\affiliation{ Department of Mathematics, Massachusetts Institute of Technology,
              Cambridge, Massachusetts 02139, USA}

\author{S.~Rotter}
\email{stefan.rotter@tuwien.ac.at}
\affiliation{ Institute for Theoretical Physics, Vienna University of
              Technology, A-1040 Vienna, Austria, EU}
\date{\today}

\begin{abstract}
  We present an efficient and flexible method for solving the
  non-linear lasing equations of the steady-state \emph{ab initio}
  laser theory. Our strategy is to solve the underlying system of
  partial differential equations directly, without the need of setting
  up a parametrized basis of constant flux states. We validate this
  approach in one-dimensional as well as in cylindrical systems, and
  demonstrate its scalability to full-vector three-dimensional
  calculations in photonic-crystal slabs. Our method paves the way for
  efficient and accurate simulations of \REFEREE{microlasers} which
  were previously inaccessible.
\end{abstract}

\maketitle

\section{Introduction}
\label{sec:intro}
As lasers become increasingly complicated, especially in nanophotonic systems with wavelength-scale features \cite{science_phc, nature_phc, englund, quantumdot}, there has been a corresponding increase in the computational difficulty of solving for their nonlinear behavior, as described by the Maxwell--Bloch (MB) equations \cite{haken_light:_1985}. To address this key challenge in the design and understanding of  lasers, a highly efficient approach to finding the non-linear steady-state properties of complex laser systems has recently been introduced, known by the acronym SALT (steady-state \emph{ab initio} laser theory) \footnote{This is the same theory as the \emph{ab initio} self consistent laser theory abbreviated as AISC, but the name has been changed.}. In this paper, we present a technique to \emph{directly} solve the SALT formulation \cite{tureci_self-consistent_2006,tureci_strong_2008,ge_steady-state_2010} of the steady-state MB equations (using finite-difference frequency-domain (FDFD) \cite{christ_fdfd, champagne_fdfd} or finite-element methods (FEM) \cite{chew_2001}), and we demonstrate that, unlike previous approaches to the SALT equations \cite{tureci_strong_2008, ge_steady-state_2010}, our technique scales to full three-dimensional (3D) low-symmetry geometries (such as photonic-crystal slabs~\cite{phc_book}).   

The SALT equations (reviewed in Sec.~\ref{sec:SALT_PDE}) simplify the general MB equations by removing the time dependence for steady-state modes, which allows SALT solvers to be potentially far more efficient than previous time-domain approaches \cite{gogny, nagra}, while providing comparable accuracy \cite{ge_quantitative_2008,cerjan_steady-state_2012}.   However, all earlier approaches to SALT required the intermediate construction of a specialized constant-flux (CF) basis for the laser modes. While efficient and yielding numerous insights in highly symmetric geometries where it can be constructed semi-analytically, the CF basis becomes unwieldy and numerically expensive for complex low-symmetry laser geometries, especially in three dimensions. In our approach, we solve the SALT equations {\it directly} as a set of coupled nonlinear partial differential equations (PDEs), using a combination of Newton-Raphson \cite{recipes}, sparse-matrix solver \cite{pastix}, and nonlinear eigenproblem \cite{guillaume} algorithms in standard FDFD or FEM discretizations.  In Sec.~\ref{sec:Examples}, we validate our solver against previous CF solutions for one-dimensional (1D) and cylindrical systems, while demonstrating that even in one dimension the CF basis  rapidly becomes large and expensive as the system is brought farther above threshold. \REFEREE{Furthermore, we show in Sec.~\ref{sec:outgoing} that analytical outgoing-radiation boundary conditions, which are difficult to generalize to three dimensions \cite{taflove}, can be substituted by the standard PML (perfectly matched layer) method \cite{taflove, champagne_fdfd, shin_fan} which is equally effective at modeling open systems}.  We also demonstrate multi-mode laser solutions (Secs.~\ref{sec:AvoidedCrossing} and \ref{sec:Examples23d}), and reproduce the non-trivial avoided crossing interaction between lasing and non-lasing modes found in Ref. \cite{liertzer_pump-induced_2012}.

We conclude in Sec.~\ref{sec:Examples23d} with full 3D vectorial laser-mode solutions for a photonic-crystal slab microcavity \cite{phc_book}. The appendixes provide further details on the computational techniques we use in this paper, but in general any standard computational method in electromagnetism could be combined with our nonlinear solver algorithms. We believe that this computational approach provides a powerful tool to design and explore laser phenomena in the complex geometries accessible to modern nanofabrication, which were previously intractable for accurate modeling.

The Maxwell-Bloch (MB) equations provide the most basic formulation of
semi-classical laser theory. The propagation of the
electromagnetic field is given by the classical Maxwell equations and
only the interaction of the field with the gain medium, represented by
an ensemble of two-level atoms embedded in a cavity or background linear medium, is treated quantum mechanically. The MB equations are a set of time-dependent
coupled nonlinear equations that are typically hard to solve
analytically, except by using many approximations and
idealizations. In the generic case of laser systems where such approximations
are not valid, the MB equations have typically been solved using
numerically expensive time-domain simulations \cite{gogny, nagra}. For the case of
steady-state lasing, as noted above, a much more efficient theory for calculating the
multi-periodic solutions of the MB equations is the \emph{steady-state ab-initio lasing theory} (SALT)
\cite{tureci_self-consistent_2006,ge_steady-state_2010,tureci_ab_2009}.
This theory has proven to be a viable tool for describing laser systems ranging from random lasers
\cite{tureci_strong_2008,stano_suppression_2013,hisch_pump-controlled_2013} to coupled laser
systems \cite{liertzer_pump-induced_2012} and photonic-crystal lasers
\cite{chua_low-threshold_2011}. It makes no {\it a priori} assumptions about
the geometry of the laser system, treats the open (non-Hermitian) character of the
laser system exactly, and the non-linear hole-burning interactions between the laser
modes to infinite order.  More realistic and quantitative laser modeling typically requires treating a gain medium with three, four or more
relevant atomic levels, but it has been shown that for the steady-state properties, under the same assumptions as SALT,
the semiclassical equations can be reduced to an effective two-level (MB) system with renormalized parameters and solved
with essentially the same efficiency as two-level SALT \cite{ge_steady-state_2010-1,cerjan_steady-state_2012}. SALT can also be used to describe
quantum properties of lasers by combining the non-linear scattering matrix of SALT with input-output theory, leading
specifically to a general formula for the linewidth of each mode in the non-linear steady-state
\cite{chong_general_2012}.

For readers familiar with linear resonant cavities in photonics, which essentially trap light for a long time in a small volume, a laser can be semiclassically understood via the introduction of nonlinear gain (amplification) whose strength is determined by an input-energy ``pump" \cite{yariv}.  As the pump strength is increased, one eventually reaches a ``threshold" at which the gain balances the cavity loss and a steady-state real-frequency lasing (``active'') mode comes into existence.  A key element is that the gain is nonlinear: increasing the laser-mode amplitude depletes the excited states of the gain medium (via a ``hole-burning" term in the gain), and so at a given pump strength above threshold there is a self-consistent stable laser amplitude. At higher pump strengths, however, this picture is complicated by the introduction of additional lasing modes, which interact nonlinearly and whose individual gains and losses are balanced simultaneously by the SALT equations. Also, while a linear ``resonant mode'' technically refers to a pole in the Green's function (or scattering matrix) at a complex frequency lying slightly below the real axis, a lasing mode can arise from \emph{any} pole that is pushed up to the real axis by the gain, even poles that start out far from the real axis and do not resemble traditional resonant-cavity modes (for example, in random lasers~\cite{tureci_strong_2008}).

%The three relevant physical quantities for a laser, which enter the MB equations are the electric field as well as the polarization and the inversion of the gain medium, all of which can depend on space and time.  By making a multi-mode steady-state ansatz and further well-motivated approximations, this system is reduced to the non-linear SALT equations for the electric field of the lasing modes.  Lasing is conceived of as the limit of an amplifying scattering process in which the input goes to zero, hence it will correspond to purely outgoing solutions with real frequency or, equivalently, to a pole in the relevant scattering matrix on the real axis. Until the external pump is strong enough for the gain to balance the loss there will be no solution of this type. However, when increasing the pump strength, non-trivial solutions appear at a sequence of thresholds and at different frequencies.  The non-linear interaction between these solutions is through the spatial hole-burning and depletion of the gain medium: each lasing mode extracts energy from the pump in a space-dependent manner which in general makes it more difficult for subsequent modes to reach threshold, and also effectively changes the index of refraction of the gain medium.

A strategy for efficiently solving the SALT equations was introduced
in \cite{tureci_strong_2008,tureci_ab_2009} and significantly extended in 
\cite{ge_steady-state_2010}. These existing methods can be viewed as a spectral integral-equation method \cite{boyd}: they solve the nonlinear problem by first parametrizing each laser mode in terms of a specialized ``spectral" basis, called the ``constant-flux (CF) states", that solve a \emph{linear} non-Hermitian Maxwell eigenproblem parametrized by its (unknown) real lasing frequency.  Because the frequency is required to be real outside the cavity, the photon flux outside the laser cavity is conserved, unlike the well-known quasi-bound states of the system, which are also purely outgoing, but do not conserve flux.  This basis is defined so that at the lasing threshold for each mode, where the non-linear hole-burning interaction term is zero, one member of the basis set {\it is} the lasing solution.  Hence, by construction, the basis expansion for the SALT solution above but near threshold converges rapidly even when the non-linear terms are taken into account, and the SALT equations reduce to finding a relatively small number of expansion coefficients for each mode.
In highly symmetric geometries such as 1D or cylindrical systems with uniform pumping, the CF states can be found semi-analytically in terms of known solutions of the Helmholtz equation in each homogeneous region (e.g., in terms of sinusoid or Bessel functions), and such a basis will typically converge exponentially quickly \cite{boyd} to the SALT solutions. Furthermore, the CF basis can be used as a starting point for other analyses of laser systems, such as to identify the cause of mode suppression due to modal interactions \cite{tureci_strong_2008,ge_steady-state_2010} and exceptional points \cite{liertzer_pump-induced_2012,brandstetter_reversing_2014}.
However, the CF basis also has some disadvantages for complex geometries or for lasers operating far above threshold where the nonlinearities are strong and the convergence is not so rapid.  In complex geometries where Helmholtz solutions are not known analytically, the CF basis itself must be found numerically by a generic discretization (e.g., FDFD or FEM) for many real frequencies (since the lasing frequency is not known {\it a priori} above threshold) and for multiple CF eigenvalues at each frequency in order to ensure convergence. The lack of separable solutions in low-symmetry two-dimensional (2D) and 3D geometries also increases the number of basis functions that are required (in contrast to cylindrical systems, for example, where the solutions $\sim e^{i m\phi}$ can be solved one $m$ at a time).   In three dimensions, where the discretization might have millions of points (e.g., on a $100\times 100 \times 100$ grid), even storing a CF basis consisting of hundreds or thousands of modes becomes a challenge, not to mention the expense of computing this many 3D eigenfunctions numerically or of computing the resulting SALT equation terms. As a consequence, our approach in this paper is to abandon the construction of the intermediate CF basis and instead to directly discretize and solve the nonlinear SALT PDEs. This approach enables us to solve even low-symmetry 3D systems, and greatly enhances the power of the SALT approach for modeling and for the design of realistic laser structures.

\section{Review of SALT}
\label{sec:SALT_PDE}
The origin of the SALT equations are the MB equations, which
nonlinearly couple an ensemble of two-level atoms with transition frequency $k_a$ $(c=1)$ to the electric field \cite{haken_light:_1985,sslbook}:
\begin{eqnarray}
\label{eq:MB1}
&- \nabla \times \nabla \times(\mathbf{E}^+)-\varepsilon_c \ddot{\mathbf{E}}^+= \frac{1}{\varepsilon_0} \ddot{\mathbf{P}}^+,& \\
&\dot{\mathbf{P}}^+ = -i(k_a-i\gamma_\bot)\mathbf{P}^+ + \frac{g^2}{i\hbar}\mathbf{E}^+D,& \\
\label{eq:MB3}
&\dot{D} = \gamma_\parallel(D_0-D)- \frac{2}{i\hbar}[\mathbf{E}^+ \cdot (\mathbf{P}^+)^*-\mathbf{P}^+ \cdot (\mathbf{E}^+)^*],&
\end{eqnarray}
Here, $\mathbf{E}^+({\bf x},t)$ and $\mathbf{P}^+({\bf x},t)$ are the positive-frequency
components of the electric field and polarization, respectively. The coupling to the negative-frequency
components is neglected in terms of a rotating wave approximation (RWA) which is both very useful for simplifying the equations and very accurate under general conditions. 
Note that at no point did we or will we assume the standard slowly-varying
envelope approximation, which, if used, reduces the accuracy of the MB solutions.
The population inversion of the medium
$D({\bf x},t)$ is given by $D_0(\mathbf{x}, d)$ in the absence of lasing, which is roughly proportional to
the external pumping rate and thus generally referred to as the pump strength.  
One of the useful features of SALT is that this pump strength can have an arbitrary spatial profile in addition to a varying global amplitude, such
that one can represent different experimental pumping protocols by evolving along a ``pump trajectory" which we parametrize here by $d$,
following Ref.~\cite{liertzer_pump-induced_2012}. Note that if there are gain atoms in unpumped regions of the laser, then the pump strength
 $D_0$ will be negative in these regions and thus the SALT equations will automatically take into account absorption due to unexcited gain atoms.
$\gamma_\bot$ and $\gamma_\parallel$ are the relaxation rates of the polarization and inversion, respectively. The linear cavity dielectric function $\varepsilon_c({\bf x})$ is homogeneous outside the cavity region, and
consequently a finite spatial domain can be used for the laser system with an outgoing boundary condition.
We have assumed a scalar $\varepsilon_c({\bf x})$ and dipole matrix element 
$g,$ % = e \langle e|{\bf x} | g \rangle$,
 although in anisotropic gain media they can be generalized to tensors. 

\REFEREE{The attractive feature of SALT is that it provides access to the spatial profiles of the lasing modes as well as to the lasing frequencies of a multi-mode microlaser at very low computational costs. To achieve this  high performance, SALT} makes two essential assumptions. First, it assumes that for a fixed pump strength
the electric field and polarization eventually reach a multi-periodic steady state,
\begin{eqnarray}
\mathbf{E}^+(\mathbf{x},t) &=& \sum_{\mu=1}^M \BPsi_\mu(\mathbf{x}) e^{-i
  k_\mu t}, \\
\mathbf{P}^+(\mathbf{x},t) &=& \sum_{\mu=1}^M {\bf p}_\mu(\mathbf{x}) e^{-i k_\mu
  t},\label{eq:saltansatz}
\end{eqnarray}
with $M$ unknown lasing modes $\BPsi_\mu$ and \emph{real} lasing frequencies $k_\mu$.
Second, SALT makes the \REFEREE{stationary inversion approximation (SIA), i.e, $\dot{D} \approx 0$. In the single-mode regime the SIA is not necessary, 
as the average inversion in steady-state is exactly zero, but in the multimode regime the inversion is in general not stationary and only under certain conditions is 
$\dot{D} \approx 0$.  However, the development of SALT was specifically oriented towards describing novel solid state 
microlasers and the necessary conditions are typically satisfied for such lasers, as we discuss in the following.}

\REFEREE{If the laser is operating in the multimode regime, then the term ${ E}(t) \cdot {P}(t)$ 
in Eq.~\eqref{eq:MB3} above will drive the inversion 
at all beat frequencies of active modes, which is of order $\Delta k$, the free spectral range of the laser.  
In addition, the polarization can respond at the rate $\gamma_\perp$ and could additionally drive time variation in the inversion.  
However, if the condition $ \Delta k, \gamma_\perp \gg \gamma_\parallel$ holds, then the inversion is being driven non-resonantly and responds quite weakly, 
except to the dc part of the drive which represents static gain saturation. The effects of the residual four-wave mixing can be 
included perturbatively if desired, as was done in Ref. \cite{ge_quantitative_2008}, but are neglected in standard SALT.  
The condition $\gamma_\perp \gg \gamma_\parallel$ is satisfied in essentially all solid state lasers due to strong dephasing, but the condition  
$ \Delta k  \gg \gamma_\parallel$ depends on the linear dimensions and geometry of the laser cavity and is typically not satisfied for 
macro scale tabletop lasers.  However for a linear cavity it typically would be satisfied for $L < 100\, \mu \rm{m}$ and hence the 
SIA tends to be a good approximation for multimode lasing in micro lasers. This general argument was made by Fu and Haken \cite{fu_multifrequency_1991} 
in 1991 and was applied to Fabry-Perot lasers, for which they provided a stability proof for the multimode state under these conditions.  
These assumptions leading to the SIA allow the derivation of the much more general SALT equations, which were then tested extensively 
in comparison to full FDTD simulations for many multimode lasing structures in Refs. \cite{ge_quantitative_2008,cerjan_steady-state_2012,liertzer_pump-induced_2012}.  A general linear stability 
analysis in the SALT framework is challenging due to the necessity of testing stability against all possible spatial fluctuations, 
something not ever done in standard analyses, where the spatial degrees of freedom are frozen.  However, work in this direction 
is in progress and partial results have been obtained.  A condition relating to the stability of multimode solutions is 
discussed immediately below.}

\REFEREE{For completeness we note that this analysis of the validity of the SIA differs from the well-established classification of 
lasers into categories denoted class A, B, and C, depending on whether two, one, or zero of the fields $E(t),P(t),D(t)$ 
can be adiabatically eliminated, meaning that the rapidly responding field instantaneously follows the slowly varying field(s). 
By far the most important case is class B, in which $P(t)$ adiabatically follows $E(t),D(t)$ (even in the transient dynamics), 
and the three MB equations are reduced to two equations for the field and the inversion.  The condition for 
class B is expressed by the inequality $\gamma_\perp \gg \gamma_\parallel, \kappa$, where $\kappa$ is the cavity decay rate.  
This condition is neither necessary nor sufficient for the validity of the SIA in the multimode regime.}

\REFEREE{The class B condition is not sufficient, as is well-known, because once two or more modes lase, the beat frequency can drive 
complex and even chaotic dynamics of the inversion and field. One needs the further condition $\Delta k \gg \gamma_\parallel$ as just noted. This laser classification was introduced before the
advent of the micro laser, for which this inequality holds, and hence it was assumed that multimode lasing would never be stable for class B.   
However, if the SIA condition $ \Delta k, \gamma_\perp \gg \gamma_\parallel$ holds then we do not need
full adiabatic elimination of $P(t)$.  The SIA and SALT can still describe the multimode steady-state which is eventually reached.  
The condition $\kappa \ll \gamma_\perp$ (``good cavity" limit) is not necessary to have a stable multimode solution. 
The magnitude of $\kappa$ only affects the steady-state in terms of its stability to fluctuations. A noise driven fluctuation will oscillate as it decays at the relaxation frequency, 
$\omega_r \sim \sqrt{\gamma_\parallel \kappa}$; if the beat frequency, $\Delta k \sim \omega_r$, then the multimode 
interference can drive the fluctuations resonantly and destabilize the multimode solution.  This mechanism was analyzed 
carefully in a number of works on the approach to chaos in lasers with injection \cite{arecchi_deterministic_1984,oppo_frequency_1986} or multiple modes. This yields a 
third implicit condition on the validity of the SIA and SALT, i.e., 
$ \Delta k \gg \sqrt{\gamma_\parallel \kappa}$. If one has a good cavity with $\kappa \ll \Delta k, \gamma_\perp$ then this is 
easily satisfied; but if one has a ``bad cavity" with $\kappa > \gamma_\perp$ (which can be achieved) then the condition can 
still be satisfied if $\gamma_\parallel$ is sufficiently small. Thus we do expect SIA
to hold in the multimode regime even for bad cavity lasers which are not standard class B, as long as this inequality holds, 
and SALT should describe multimode lasing in the bad-cavity limit. Comparisons of SALT with FDTD for bad cavities confirm this 
expectation, as well as recent work by Pillay \emph{et al}.~which uses SALT to compute the laser linewidth in the bad cavity limit \cite{pillay}.}

Using these well-motivated approximations, Eq.~\eqref{eq:MB1} can then be written for each lasing mode $\BPsi_\mu (\mathbf{x})$ as
\begin{equation}
  \label{eq:SALTsystem}
	\Big[ -\nabla \times \nabla \times  + \, k^2_\mu \varepsilon_c(\mathbf{x}) + k^2_\mu \gamma(k_\mu) D \Big]
	\BPsi_\mu ({\bf x}) = 0,
\end{equation}
where the two-level active gain material is described by the
non-linear susceptibility $\gamma(k_\mu) D$. Here, $\gamma(k_\mu)$,
is the Lorentzian gain curve, where
\begin{equation}
\label{eq:gamma}
  \gamma(k_\mu) \equiv \frac{\gamma_\bot}{k_\mu - k_a + i \gamma_\bot},
\end{equation}
and $D$ the population inversion. The latter contains the spatial hole-burning term that nonlinearly couples all lasing modes,
\begin{equation}
  \label{eq:NLcoupling}
  D({\bf x},d,\{k_\nu, \BPsi_\nu\}) =  \frac{D_0({\bf x}, d)} {1 +
  \sum_{\nu=1}^M \left| \gamma(k_\nu) \BPsi_\nu(\mathbf{x}) \right| ^2},
\end{equation}
where the $\BPsi_\nu(\mathbf{x})$ are in their natural unit ${e_c = 2 g / \hbar \sqrt{\gamma_\perp \gamma_\parallel}}$.
%The three relevant physical quantities for a laser, which enter the MB equations are the electric field as well as the polarization and the inversion of the gain medium, all of which can depend on space and time.  By making a multi-mode steady-state ansatz and further well-motivated approximations, this system is reduced to the 

The non-linear SALT equations,  Eq.~(\ref{eq:SALTsystem}), for the electric field of the lasing modes, $\BPsi_\mu ({\bf x})$,  and for the associated lasing frequencies $k_\mu$
%.  Lasing is 
can be conceived of as the limit of an amplifying scattering process in which the input goes to zero, corresponding to purely outgoing solutions with real frequency or, equivalently, to a pole in the relevant scattering matrix on the real axis. Until the external pump is strong enough for the gain to balance the loss there will be no solution of this type, i.e., $\BPsi_\mu ({\bf x})=0$. However, when increasing the pump strength, non-trivial solutions appear at a sequence of thresholds and at different frequencies $k_\mu$.  The non-linear interaction between these solutions is through the spatial hole-burning and depletion of the gain medium, Eq.(\ref{eq:NLcoupling}): each lasing mode extracts energy from the pump in a space-dependent manner which in general makes it more difficult for subsequent modes to reach threshold, and also effectively changes the index of refraction of the gain medium.
%It is  interesting to point out here a similarity to the set of N coupled nonlinear PDEs that describe the physics of incoherent vector solitons, see Appendix \ref{app:vector_solitons}.

As already noted, Eq.~\eqref{eq:SALTsystem} has been solved in 1D and 2D geometries, where either the electric or the magnetic field can 
be treated as scalar, for diverse systems such as random, microdisk or photonic crystal lasers using an algorithm based on expansion of the solutions in the CF basis \cite{ge_steady-state_2010}.
In the most recent and most efficient formulation, the \emph{linear} non-Hermitian eigenvalue problem,
\begin{equation}
\label{eq:CF}
\left[ -\nabla \times \nabla \times + \, k^2 \varepsilon_c(\mathbf{x}) + k^2 \eta_n(k) f(\mathbf{x}) \right] u_n (\mathbf{x};k) = 0,
\end{equation}
is used to define the optimal set of threshold CF states $u_n(\mathbf{x}; k)$ and eigenvalues $\eta_n(k)$.

The function $f(\mathbf{x})$ adapts the basis to the
spatial pump profile of the experiment of interest and is nonzero only inside the gain medium. The $u_n(\mathbf{x}; k)$ form a complete basis and satisfy a biorthogonality relation at any frequency $k$. Equation \eqref{eq:SALTsystem} is solved by projecting the lasing modes $\Psi_\mu(\mathbf{x})$ into the CF basis. The resulting non-linear eigenvalue equation can only be satisfied at discrete frequencies which hence determine the lasing frequencies, $k_\mu$.
In principle one does not need to pre-calculate and store the CF basis at different real values of $k$ but it is numerically favorable to do so in general. However, the wider the Lorentzian gain curve, Eq.~(\ref{eq:gamma}), is compared to the free spectral range, 
the more memory intensive the storage 
of the CF basis becomes, which makes calculations problematic in two and three dimensions.
Moreover, if the pump profile $f(\mathbf{x})$ is fixed and only its amplitude is varied experimentally, then CF states need only be calculated for various $k$ values, but
if the pump profile also varies along a pump trajectory then one has to calculate new CF states also for many values of $d$ \cite{liertzer_pump-induced_2012}.
For a limited set of highly symmetric cavities, including piecewise-homogeneous 1D slabs and uniform cylinders, the solution of Eq.~\eqref{eq:CF} is known semi-analytically at any $k$. However, for all other geometries, Eq.~\eqref{eq:CF} must be solved numerically for all relevant $k$ needed to build a basis. Consequently, for a fully-vectorial treatment of SALT in arbitrary cavities, CF bases cannot be used without significant computational costs. Our direct solution method eliminates the computation and storage of CF bases and scales easily to 3D geometries.

\section{Solution method}
\label{sec:SIM}

	\subsection{Overview}
	\label{sec:solution_strategy}
	The basic idea of our new solution method to obtain the lasing modes in the SALT is as follows: We discretize Eq.~\eqref{eq:SALTsystem}, 
using standard discretization techniques like FEM (see Appendix \ref{app:FEM}) or FDFD (see Appendix \ref{app:FDFD}),
and iteratively solve for the lasing modes $\BPsi_\mu$ and their frequencies $k_\mu$ at successively increasing values of the pump parameter $d$. This nonlinear coupled problem is most conveniently solved by using the Newton-Raphson method.
For initial guesses, we use the modes at threshold when we are close above threshold, and the modes at the previous pump step when we are far above threshold. 
In order to find the first threshold and the corresponding solution, Eq.~\eqref{eq:SALTsystem} is initially solved for $d=0$ as an eigenvalue problem (EVP). 
The solutions are the resonances or quasi bound states $\barpsi_n$ of the passive cavity, corresponding to the poles of the passive scattering matrix ($S$ matrix) \cite{ge_steady-state_2010} with frequencies $\bark_n$ lying in the negative imaginary half plane 
(note that we will label all quantities below threshold with
overbars throughout the paper). 
While increasing the pump $d$, Eq.~\eqref{eq:SALTsystem} is solved without the nonlinearity in Eq.~\eqref{eq:NLcoupling} and the nonlasing modes near the gain frequency $k_a$ are tracked until the first $\bar{k}_{n_0}$ reaches the real axis and turns the corresponding mode into an active 
lasing mode, $\barpsi_{n_0}\to\BPsi_1$. 
Once we have crossed the first threshold, we use the solutions for $\barpsi_{n_0}$ and $\bark_{n_0}$ of the eigenvalue problem at threshold as a first guess for the solution of $\BPsi_1$ and $k_1$ in the non-linear Newton solver slightly above threshold. The latter already includes the non-linearity $D(\mathbf{x}, d, \{k_1, \BPsi_1\})$ which, once the Newton solver has converged, we treat as a fixed function like $\varepsilon_c(\mathbf{x})$ to examine the remaining non-lasing modes $\barpsi_n$ at the current pump strength $d$. 
This has to be done in order to verify if further modes cross the lasing threshold.
For the non-lasing modes, Eq.~\eqref{eq:SALTsystem} is thus only nonlinear in $\bark_n$ and linear in $\barpsi_n$, 
such that this problem can be cast into a nonlinear EVP~\cite{guillaume}. %, as explained in Sec.~\ref{sec:inactive_modes}.\\
The procedure of increasing the pump is now continued by tracking the lasing mode  solving the nonlinear coupled SALT system, while the non-lasing modes are evaluated from the corresponding nonlinear EVP until a second mode reaches threshold. At this point the number of lasing modes is increased by 1 and the procedure continues with two and more lasing modes in essentially the same way. 

\begin{figure}[tb]
  \includegraphics{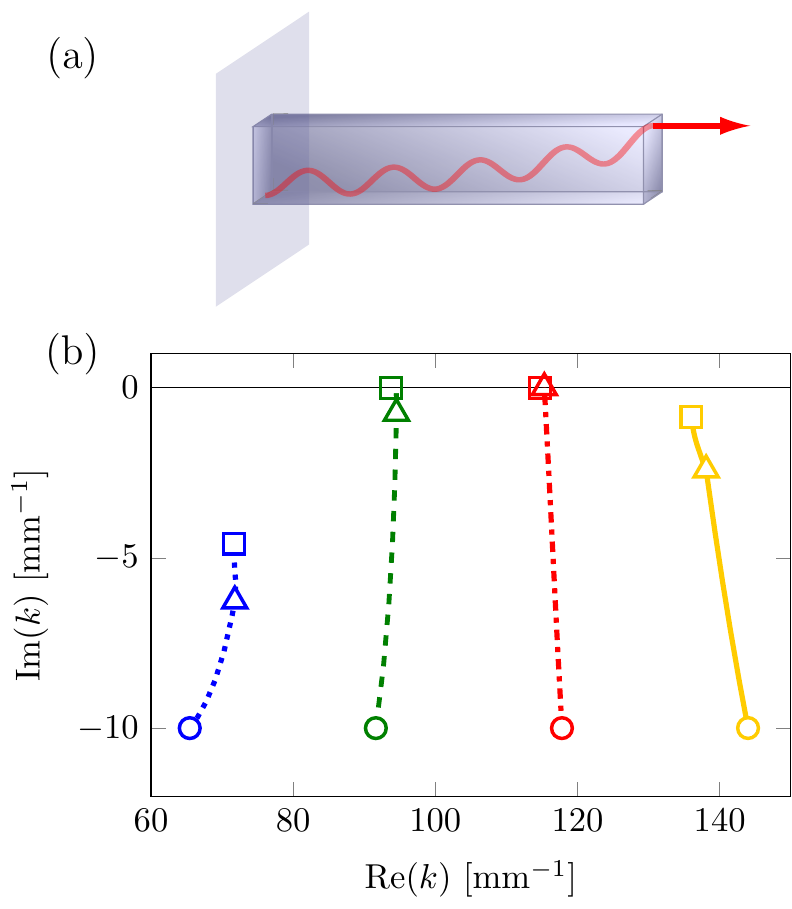}
        \caption{(Color online) (a) 1D slab cavity laser of length $L=100\ \mu\mathrm{m}$ with purely reflecting boundary on the left side and open boundary on the right side. The
        mode shown in red (gray)corresponds to the intensity profile of the first lasing mode at threshold.
        (b) SALT eigenvalues corresponding to the scattering matrix poles for a uniform  and linearly increasing pump strength $D_0(\mathbf{x}, d) = d$
         applied inside the slab [$D_0(\mathbf{x}, d) = 0$ outside]. We use a refractive index 
         $\sqrt{\varepsilon_c}=1.2$ in the slab ($\sqrt{\varepsilon_c}=1$ outside), a
         gain frequency ${k_a = 100\ \mathrm{mm}^{-1}}$ and a polarization relaxation rate $\gamma_\bot = 40\ \mathrm{mm}^{-1}$. The trajectories 
         start at $d=0$ (circles) and move toward the real axis with different speed when increasing $d$. The first lasing mode (dash-dotted red line) 
         activates at $d=0.267$ (triangles) with $k_1 = 115.3\ \mathrm{mm}^{-1}$.  The trajectories end at $d=1$ (squares) where a second lasing mode (dashed green line) 
         turns active and the two other non-lasing modes (blue dotted and yellow solid line) remain inactive. The values at $d=1$ coincide 
         with the data in Fig.~\ref{fig:iSALT}.  
}
  \label{fig:linEVP}
\end{figure}

To illustrate this approach in more detail, we apply it to the simple one-dimensional edge-emitting laser shown in Fig.~\ref{fig:linEVP}(a) which 
already captures all the main features. We pump the 1D slab cavity uniformly along its length $L=100\,{\rm \mu m}$ with a pump strength $D_0({\bf x},d) = d$ which,
above the first threshold, leads to emission to the right. Starting with $d=0$, 
where the  SALT system reduces to a simple resonance problem, we increase $d$ and observe that the resonance poles move upwards in the complex plane; 
see Fig.~\ref{fig:linEVP}(b) where the starting point $d=0$ is marked by circles and the pump value at the first threshold, ${d_1 = 0.267}$, 
is marked by triangles.
Below this first threshold no mode is lasing, such that the non-linear spatial hole-burning term is zero, resulting in the following PDE 
for all non-lasing modes,
\begin{equation}
 \label{eq:linSALT}
 \left\lbrace- \nabla \times \nabla \times+ \bark_n^2 \big[\varepsilon_c({\bf x}) + \gamma(\bark_n) D_0({\bf x},d) \big]\right\rbrace \barpsi_n({\bf x})  =  0\,,
\end{equation}
which is linear with respect to $\barpsi_n$, but into which the resonance values $\bark_n$ enter non-linearly. 
Starting at the first threshold, the terms $\BPsi_1$ and $ k_1$ of the first lasing mode enter the spatial 
hole-burning denominator in Eq.~\eqref{eq:NLcoupling} 
(where $M=1$),  
resulting in the following equation for the first lasing mode $\BPsi_1$ and its wavenumber $k_1$,
\begin{equation}
 \label{eq:one_mode_SALT}
 \Big\lbrace - \nabla \times \nabla \times+ k_1^2 \Big[\varepsilon_c({\bf x}) + \ \REFEREE{\frac{\gamma(k_1) D_0({\bf x},d)}{1+  |\gamma(k_1)\BPsi_1({\bf x})|^2} } \Big]  \Big\rbrace \BPsi_1({\bf x})  =  0 
\end{equation}
which is now nonlinear with respect to both $\BPsi_1$ and $k_1$.
When continuing to increase the pump, the frequencies corresponding to
the active modes are forced to stick to the real axis, while the eigenvalues
associated to all other inactive modes continue moving upwards, see Fig.~\ref{fig:linEVP}(b).  
To detect the activation of further modes, the inactive modes have to be
recalculated again, however, this time by additionally taking into
account the spatial hole burning contribution of the currently lasing
active mode $(\BPsi_1,k_1)$ at a given pump strength $d$. For this, we insert the currently
active mode into the denominator of Eq.~\eqref{eq:one_mode_SALT} which
turns the above nonlinear problem into another nonlinear (in $\bark_n$) eigenvalue problem,
\begin{equation}
\Big\{- \nabla \times \nabla \times+ \bark_n^2\Big[ \varepsilon_c({\bf x}) + \REFEREE{\frac{\gamma(\bark_n)D_0({\bf x},d)}{1+  |\gamma(k_1) \BPsi_1({\bf x})|^2}}\Big] \Big\} \barpsi_n({\bf x})  =  0 \label{eq:inactivemodes}
\end{equation}
which, however, has the same structure as Eq.~\eqref{eq:linSALT}. As soon as the
imaginary part of another eigenvalue $\bark_n$ reaches the real axis, a new
laser mode $\BPsi_2$ becomes active which increases the size of the
nonlinear problem by 1. For even higher pump strength and a larger number of lasing modes this procedure
continues accordingly. Note also, that the case when a mode shuts down during the pumping process can be 
incorporated without major effort. 

To summarize, the solution of the SALT equations reduces essentially to
computing the full nonlinear (in $\BPsi_\mu$ and $k_\mu$) system of PDEs through
a Newton-Raphson method and the computation of an EVP  which is linear in $\barpsi_n$ %by fixing the current spatial hole burning effects, 
but which still remains nonlinear in $\bark_n$.  Details of how to obtain the \emph{active} or \emph{lasing} 
solutions $\left\lbrace\BPsi_\mu,k_\mu\right\rbrace$ of the Newton problem as well as the \emph{inactive} or \emph{non-lasing} solutions
$\left\lbrace\barpsi_n,\bark_n\right\rbrace$ through the 
nonlinear EVP are provided in the following two sections.

	\subsection{Lasing modes}
	\label{sec:active_modes}
	For modes that are lasing, Eq.~\eqref{eq:SALTsystem} is nonlinear in the 
unknowns $\{ \BPsi_\mu(\mathbf{x}), k_\mu \}$. 
As theses modes are all coupled together through the spatial hole-burning interaction, they must be solved simultaneously.
In general, such systems of nonlinear equations can be written in the form 
 \begin{equation}
 \label{eq:abstractF} 
 \mathbf{f}(\mathbf{v})=0
% F(\{ \BPsi_\mu(\mathbf{x}), k_\mu \})=0
 \end{equation}
where the vector of equations $\mathbf{f}$ is an analytic nonlinear function of the unknown solution vector $\mathbf{v}$ which again gathers all unknowns $\{ \BPsi_\mu(\mathbf{x}), k_\mu \}$.
This nonlinear problem can generally be solved by using the Newton-Raphson method~\cite{recipes}.
The basic idea is that for a guess $\mathbf{v}_0$ for the solution~$\mathbf{v}$, one can write
\begin{equation}
\label{eq:newton}
\mathbf{v}-\mathbf{v}_0=-\mathcal{J}(\mathbf{v}_0)^{-1}\mathbf{f}(\mathbf{v}_0) + \mathcal{O}(\left|\mathbf{v}-\mathbf{v}_{0}\right|^2),
\end{equation}
 where $\mathcal{J}$ is the Jacobian matrix of partial derivatives of $\mathbf{f}$ with respect to $\mathbf{v}_0$. A solution $\mathbf{v}$ can usually be obtained by iterating Eq.~\eqref{eq:newton} using only the linear terms. This iterative algorithm converges ``quadratically" (squaring the errors on each step \cite{recipes}) if $\left| \mathbf{v}_0 - \mathbf{v} \right|$ is small. 
Further, we use an analytic evaluation for the Jacobian $\mathcal{J}$ from Eq.~\eqref{eq:SALTsystem}, as described in Appendix~\ref{app:jacobian},  and do not need to compute it using numerical differentiation schemes. Since $\mathcal{J}$ is then a sparse matrix 
each iteration can exploit fast algorithms for sparse linear equations \cite{trefethen, davis_sparse}.

To solve Eq.~\eqref{eq:abstractF} on a discrete level, 
we project the complex fields $\BPsi_\mu (\mathbf{x})$ of each lasing mode onto a discrete $N$-component basis as explained in Appendixes~\ref{app:FEM} (for a FEM approach) and \ref{app:FDFD} (for an FDFD approach).
 Unlike the CF basis, we use a \emph{localized} basis generated once from a grid or mesh. This is the key to producing \emph{sparse} matrices and hence makes the method scalable to the larger bases required in two and three dimensions. 
The discretizations on such a basis 
turn the fields $\BPsi_\mu$ into complex coefficient vectors $\mathbf{c}_\mu$, while $k_\mu$ is required to be purely real.
Because the SALT equations are not differentiable in the complex fields (due to the complex conjugation), 
we split  our unknown coefficient vectors $\mathbf{c}_\mu$ (and the vector function $\mathbf{f}$ accordingly) into their real and imaginary parts.
The discretized version of $\mathbf{v}$ then consists of
$(2N+1)M$ real unknowns (fields and frequencies). However, we only obtain $2NM$ real-valued equations from $\mathbf{f}$. The underspecification comes from the fact that the hole-burning term $D(\mathbf{x}, \{k_{\nu}, \BPsi_{\nu} \} )$ happens to be invariant under global phase rotations
$\BPsi_{\nu} (\mathbf{x}) \rightarrow e^{i \phi_{\nu} } \BPsi_{\nu} (\mathbf{x})$. In addition to the problem of underspecification, there is also a problem of stability: for lasing modes slightly above threshold, the amplitude is nearly zero, which would result in problems distinguishing between the solution we want and the trivial solution $\BPsi (\mathbf{x})=0$. We resolve both issues by normalizing the amplitude and fixing the phase of all lasing modes while keeping track of their amplitudes using a separate variable. This procedure results in both the number of real unknowns and the number of real equations being $(2N+2)M$. Further details for our method for lasing modes are given in Appendix~\ref{app:jacobian}.

Note that for the Newton-Raphson iteration to be scalable to higher dimensions and to high-resolution meshes, it is also important to use
a scalable solver  (in our case, the sparse direct solver \cite{davis_sparse} PaStiX \cite{pastix} was called from the PETSc library \cite{petsc-efficient} because the Jacobian is sparse). 
For very large-scale 3D systems, it may become necessary to use iterative linear solvers \cite{trefethen}  for each Newton step instead, in which case
it is important to select certain PML formulations \cite{shin_fan}.

	\subsection{Non-lasing modes}
	\label{sec:inactive_modes}
	In order to find the first pump threshold and the corresponding lasing solution as well as to verify when a new mode activates, the non-lasing modes 
have to be monitored while changing the pump. 
These non-lasing modes $\barpsi_n$ are defined as complex-frequency solutions to Eq.~\eqref{eq:SALTsystem} that do not enter into the nonlinear
hole burning term in $D(\mathbf{x}, d, \{k_\nu, \BPsi_\nu\})$, see Eq.~\eqref{eq:NLcoupling}. 
Due to causality constraints, the complex eigenvalues associated with non-lasing modes, 
$\bark_n$ always feature $\Im (\bark_n) < 0$, and usually approach the real axis as $d$ is increased (interesting exceptions are discussed in 
Sec.~\ref{sec:AvoidedCrossing}).
When all lasing modes have been determined for a particular $d$, the function $D(\mathbf{x}, d, \{k_\nu, \BPsi_\nu\})$ is known and can be treated as a fixed function like $\varepsilon_c (\mathbf{x})$, see Eq.~(\ref{eq:inactivemodes}). 
As outlined in Sec.~\ref{sec:solution_strategy}, this reduces Eq.~\eqref{eq:SALTsystem} 
to a non-Hermitian, nonlinear eigenvalue problem (NEVP) which is \emph{linear} in the eigenvectors $\barpsi_n (\mathbf{x})$, but nonlinear in the 
complex eigenvalues $\bark_n$.

For situations where we are only interested in the behavior of a few lasing modes in a small range of the pump parameter $d$,
Newton's method is still a convenient approach to determine the non-lasing modes
and, in fact, the only viable method in terms of computational cost for high resolution 2D or 3D computations.
In this case, we typically use standard EVP algorithms to solve Eq.~\eqref{eq:SALTsystem} first for $d = 0$ 
(which is usually either linear or quadratic in $\bark$, depending on the method for implementing the outgoing radiation condition). 
This provides us all the modes of interest which we then track 
to threshold with Newton's method as $d$ is increased. 
As in Sec.~\ref{sec:active_modes}, convergence is ``quadratic", but, unlike for the lasing modes, Eq.~\eqref{eq:newton}
can be used with complex unknowns and equations since Eq.~\eqref{eq:SALTsystem} is differentiable in 
all unknowns once $D(\mathbf{x}, d, \{k_\nu, \BPsi_\nu\})$ is fixed. 
The downside of Newton's method is that, in the absence of a good initial guess, it can be very unpredictable 
and slow to converge. Such a situation arises, e.g., when the modes that can lase are not known {\it a priori} 
as in the case where a large number of near-threshold modes are
clustered together, all with frequencies close to the gain center $k_a$. In this instance, a more general and comprehensive method for
evaluating the non-lasing modes is required.  

Such more general techniques exist in terms of NEVP solvers \cite{guillaume}.
One conceptually simple method for our problem is to divide Eqs.~(\ref{eq:SALTsystem}) and \ref{eq:inactivemodes}) by $\gamma(k)$, 
turning the rational EVP into a cubic EVP which can then be linearized at the expense of making the problem three times as large 
and possibly also very ill-conditioned. Other, more sophisticated solution 
methods include ``trimmed" linearization \cite{su_solving_2011}, Newton \cite{newton_inverse}, Jacobi-Davidson \cite{jacobi_davidson}, rational Krylov
\cite{rational_krylov}, and nonlinear Arnoldi~\cite{nonlinear_arnoldi}. 
Independently of the chosen solution strategy, we can take into account  that 
only modes which have a spectral overlap with the gain curve $\gamma(k)$ near its center frequency $k_a$ 
are expected to be candidates for active laser modes. In addition, the Lorentzian gain curve of width $\gamma_\bot$  produces a singularity 
in the NEVP at $k=k_a -i\gamma_\bot$ which may result in spurious numerical solutions.
Combining these observations, we restrict our attention to those eigenvalues $\bark_n$ that are in the following cropped subpart
of the complex plane:
 $\{z\in \mathbb{C} \,|\,\Im(z) > -\gamma_\bot\wedge\Re(z)\in [k_a -\gamma_\bot,k_a +\gamma_\bot]\}$.  
A suitable method that allows us to conveniently include such auxiliary restrictions is the contour integral method
presented recently in \cite{asakura_sakurai-contour-2010,beyn_integral_2012} and reviewed in Appendix~\ref{app:beyn}.
There, the search for eigenvalues is restricted to a region within a smooth contour such as a
circle or an ellipse; see Fig.~\ref{fig:iSALT}(a). By using the residue theorem, all poles of the inverse of the
differential operator, which are equivalent to the eigenvalues of the
same operator, are obtained within the specified contour. This feature is not only useful for employing this method as a stand-alone solver for
non-lasing modes, but also as a complementary tool to check if, in addition to the limited set of non-lasing modes that are tracked with 
a Newton solver, no new modes have entered the region of interest within the chosen contour.

	\subsection{Alternative strategy for a single pump}
	\begin{figure}[!h]
  \includegraphics[width=0.9\columnwidth]{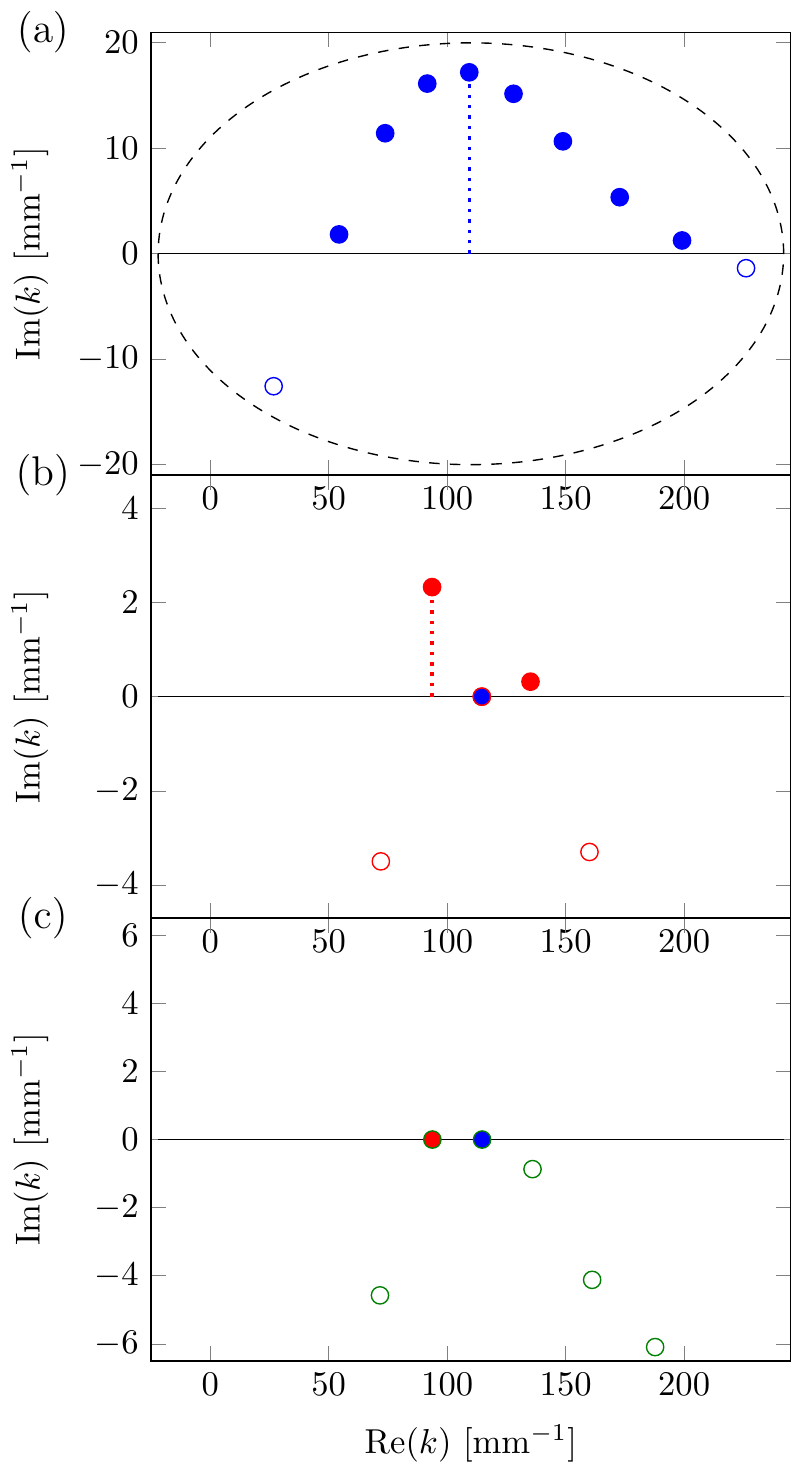}
    \caption{(Color online) Single pump algorithm applied to the 1D edge emitting laser
    shown in Fig.~\ref{fig:linEVP}(a). Here the lasing modes for the single pump strength $ D_0 = 1$ are obtained 
    within three iteration steps (see blue, red, and green colors, respectively). (a) In the first step, the eigenfrequencies of 
    Eq.~\eqref{eq:linSALT}  are determined for  $ D_0 = 1$. Full and empty circles represent modes in the upper (non physical) and lower (nonlasing) part    
    of the complex plane, respectively. The dashed ellipse indicates the boundary inside of which all eigenvalues are determined using the contour 
    integral method. % in Sec.~\ref{sec:inactive_modes}.
    The dotted vertical line marks the most non physical mode which is used as a first guess of a lasing mode in the next step (b).
    This ansatz shifts not only the corresponding eigenvalue down to the real axis, but also the
    other eigenvalues are shifted downwards due to the resulting pump depletion (see modes indicated in red).   
    (c) After including again the most non-physical mode of the previous step as a guess for the second lasing mode (see red dot) and performing 
    the corresponding iteration with Newton, we obtain the solution which coincides exactly 
    with the data in Fig.~\ref{fig:linEVP}(b) (see squares there), where two modes are active while all other modes are non-lasing.}
  \label{fig:iSALT}
\end{figure}

Similar to the single-pole approximation in the CF-expansion method  \cite{ge_steady-state_2010}, it
is possible to speed up the calculations of the direct solver 
when the intensity of the laser is only desired at or starting from a
specific pump strength $d_0$.
For this, we first solve the SALT equations, Eq.~\eqref{eq:SALTsystem}, only at this
desired pump strength $d_0$ by neglecting any spatial hole-burning interactions in 
$D(\mathbf{x}, d, \{k_\nu, \BPsi_\nu\})$. If $d_0$ happens to be above the first threshold, the corresponding NEVP will yield 
complex frequencies $k_n$ that partly lie in the non physical region above the real axis in the complex plane. This is shown in
Fig.~\ref{fig:iSALT}(a), again for the simple 1D edge-emitting
laser introduced above. Next, the  most  non-physical mode, i.e., the one which has the eigenvalue with 
the highest imaginary part, is selected and the corresponding solution vector as well as the real part of the corresponding eigenvalue are
used as initial guesses in
the nonlinear SALT solver. After the nonlinear iteration converges, the
corresponding solution is then included in the spatial hole burning
term, which effectively reduces the pump within the system and pulls down all
inactive modes in the complex plane; see
Fig.~\ref{fig:iSALT}(b). If some of the remaining inactive modes are
still located above the real axis, this procedure is repeated
by increasing the number of active lasing modes until all modes lie on
or below the real axis. The latter are then the true lasing modes of the
SALT at the desired pump strength $d_0$; see Fig.~\ref{fig:iSALT}(c).
Hence, as long as the nonlinear solver manages to converge, a solution
to the SALT can be obtained rather quickly.

	\subsection{Outgoing radiation condition}
	\label{sec:outgoing}
	For numerical computations, the outgoing radiation condition must be implemented within a truncated, finite domain. 
In one dimension, the radiation condition can be expressed exactly \cite{tolstikhin_siegert_1998}. This also allows us to shift the 
boundary of the domain right to the border of the cavity, which decreases the computational cost. 
This method is, however, not easily generalizable to two and three dimensions \cite{taflove}. 
\begin{figure}[tb]
  \includegraphics[width=\columnwidth]{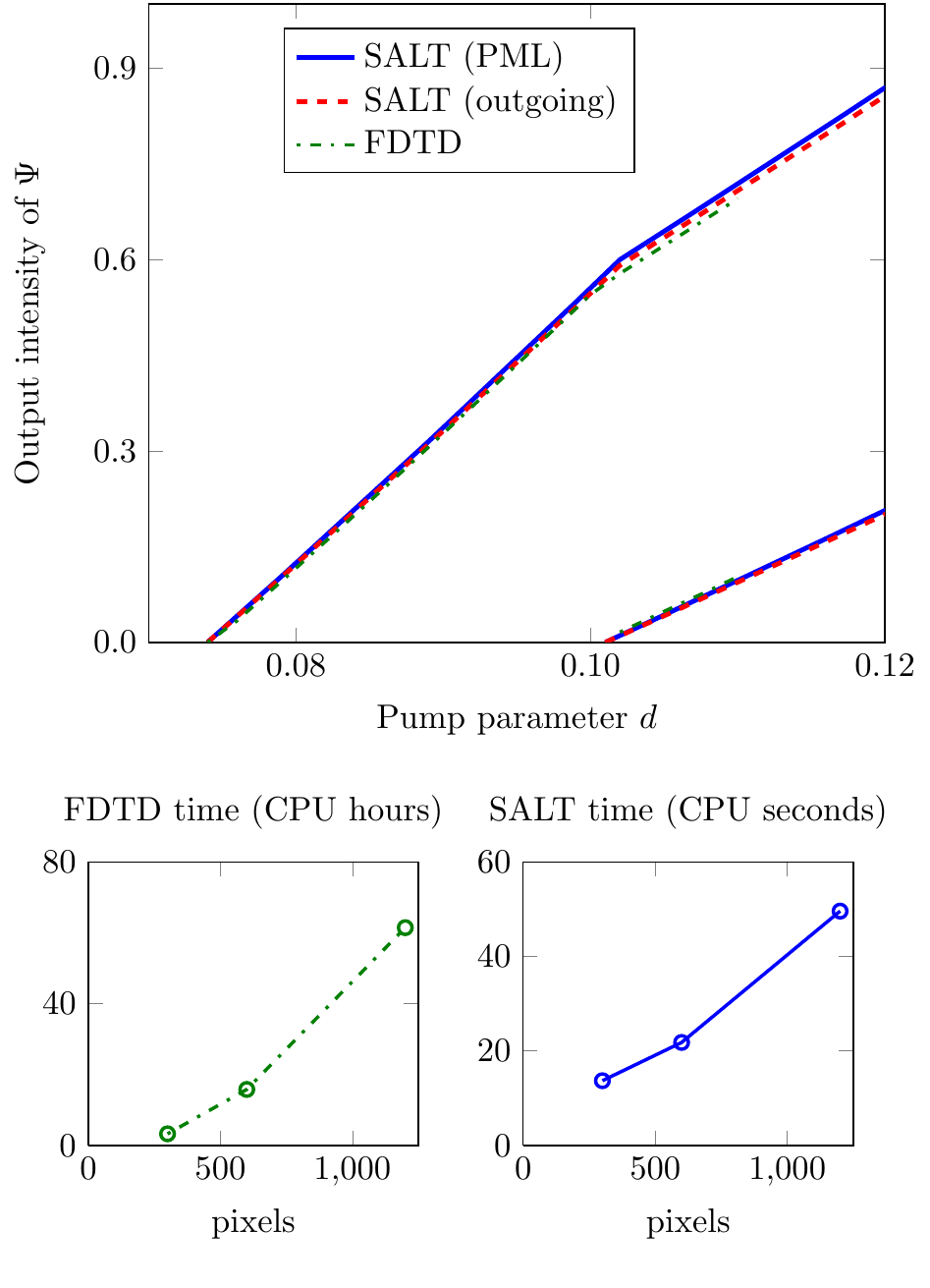}
  \caption{(Color online) Comparison between the laser output using SALT with
  exact outgoing boundary conditions and PML absorbing layers, on the one hand, and a full time integration of the MB equations using FDTD, on the other hand.  
  We study the \REFEREE{first and second TM lasing modes} of a 1D slab cavity which is similar to the one above. The applied pump is uniform, $D_0({\bf x},d)=d$, 
   the cavity has a uniform dielectric $\sqrt{\varepsilon_c} = 2$ a 
   length $L = 100\ \mu\mathrm{m}$, and gain parameters $\gamma_{\perp} = 3 \ \mathrm{mm}^{-1}, k_a = 300\ \mathrm{mm}^{-1}$. For the FDTD simulations additionally $\gamma_\parallel = 0.001\ \mathrm{mm}^{-1}$ was used. The PML method
   is nearly as accurate as the outgoing boundary condition, but has the advantage of being easily generalizable to two and three-dimensional calculations \cite{taflove}. 
   \REFEREE{The times to reach $d=0.11$ are shown for the two methods (with identical spatial resolution). The FDTD computation was done on 
   the Yale BulldogK cluster with E5410 Intel Xeon CPUs, while the SALT computations were done on a Macbook Air.}
    }
  \label{fig:outgoing_pml}
\end{figure}
An efficient and robust alternative is to use the standard perfectly matched layer (PML) technique \cite{MR1294924,MR2353940} in which an artificial material is placed at the boundaries. The material has a certain complex permittivity and permeability such that it is absorbing and analytically reflectionless.
In one dimension, the PML technique can be tested against an exact outgoing boundary condition, and the two methods yield results that are 
nearly indistinguishable, as shown in \fig{outgoing_pml}. 
Also shown in \fig{outgoing_pml}  is a comparison with conventional methods of solving the MB equations using finite difference time domain (FDTD) simulations demonstrating the validity of the stationary inversion approximation used in the derivation of the SALT equations. Both the quantitative agreement between SALT and FDTD solutions as well as the former's substantial numerical efficiency over the latter have been previously documented \cite{ge_quantitative_2008,cerjan_steady-state_2012}. \REFEREE{Of course, the precise computation times depend on many factors, including hardware details, parameter choices in the algorithms, and software implementation quality, but the magnitude of the difference here makes it unlikely that any FDTD implementation could be competitive with the SALT approach.}

\section{Assessment and application of the solution method}
\label{sec:Examples}
In this section we will validate our solution strategy against the traditional method based on CF states and 
we will show first results for prototypical laser cavities.

	\subsection{1D slab laser as test case}
	\label{sec:Examples1d}
	We demonstrate here the accuracy of the presented direct solver method
by studying in more detail the 1D edge-emitting slab laser introduced in Sec.~\ref{sec:solution_strategy}.  One of
the advantages of the direct solver, as compared to the CF method, is
\begin{figure}[tb]
  \includegraphics{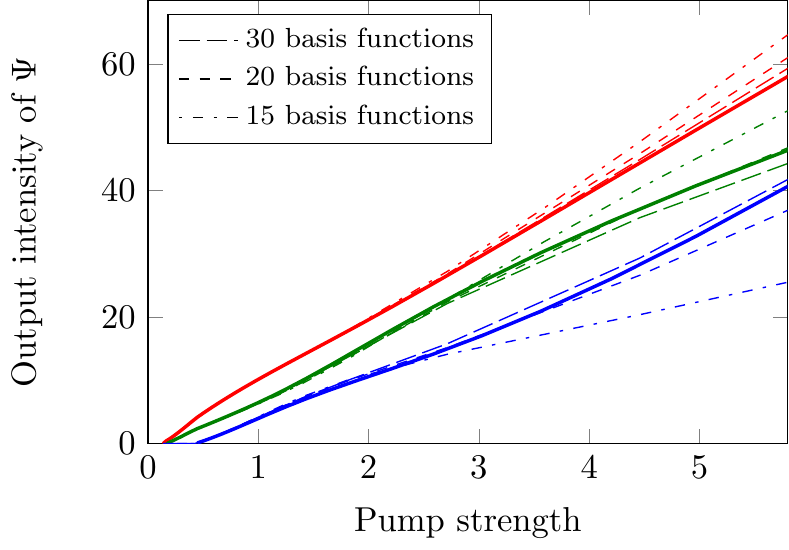}
  \caption{(Color online) Output intensity vs pump strength in a 1D resonator with
    reflecting boundary on the left side and outgoing radiation on the
    right side; see Fig.~\ref{fig:linEVP}(a).  
    The cavity has length $100\ \mu\mathrm{m}$ with a refractive
    index $n = 1.01$. The gain curve has its peak at $k_a = 250\ \mathrm{mm}^{-1}$ and a
    width $ 2\gamma_\bot = 15\ \mathrm{mm}^{-1}$.  The output intensity is given by $|\BPsi|^2$ evaluated at the right boundary $x=L$. The pump is  constant in
    the entire cavity.  Solid lines describe the results of our
    solution method. Comparing them to the solutions of the
    CF-state formalism with 30 (long dashed), 20 (dashed), and 15
    (dash-dotted) CF-basis functions, one observes that the two approaches converge towards each other for a sufficiently high number of CF states
    being included.
    }
  \label{fig:lowQ_basis_comp}
\end{figure}
the accuracy of its solutions far above the threshold. 
In this regime the CF basis becomes a poorer match
for the lasing modes and, as explained in Sec.~\ref{sec:intro}, a large number $N_\mathrm{CF}$ of basis functions is required for
convergence compared to near threshold. This is especially relevant for
low-$Q$ (short-lifetime) laser resonators such as random lasers or cavities featuring gain-induced
states, as considered, e.g., in \cite{ge_unconventional_2011} .  In Fig.~\ref{fig:lowQ_basis_comp} the intensity of such a
low-$Q$ cavity is plotted with respect to an overall pump strength $d$
for a constant spatial pump profile. The figure contains both the
results of the direct and of the CF state solver. For the latter the solution
for different numbers $N_{\mathrm{CF}}$ of CF states are depicted, demonstrating 
that for a larger basis the solution converges towards the
solution of the direct solver. Our solution method thus leads to
an accuracy far above threshold which can only be achieved by the
traditional approach with a considerably large number of CF states.

	\subsection{Non-uniform pump and avoided resonance crossings}
	\label{sec:AvoidedCrossing}
\begin{figure}[tb]
  \includegraphics[width=\columnwidth]{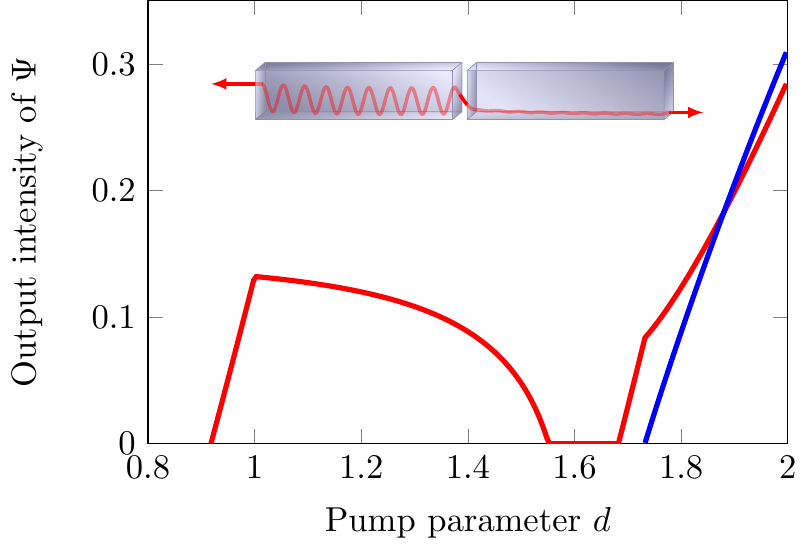}
  \caption{(Color online) Output intensity versus pump strength in a laser system of
    two 1D cavities, each of length $100\ \mu\mathrm{m}$ and an air gap of size
    $10\ \mu\mathrm{m}$, see inset.  The refractive index of the cavity material is $n =
    3+0.13i$ and the gain curve is centered at $k_a = 94.6\ \mathrm{mm}^{-1}$ with a
    width of $2\gamma_\bot = 4\ \mathrm{mm}^{-1}$. For the pump parameter in the interval $0<d<1$ 
    the pump is linearly increased in the left resonator from zero to $D_{\text{max}} = 1.2$ (the intensity pattern of the mode lasing at $d=1$ is shown in the inset). 
    For $1<d<2$ the pump in the left resonator
    is kept at the value of $d=D_{\text{max}}$ and the pump in the right resonator is increased from zero to the same value as
    on the left.  The output intensity here is given by the sum of $|\BPsi|^2$ evaluated on both open ends. As a result of this pump-trajectory, a non-monotonous evolution of the total emitted laser light intensity is observed
    with a complete laser turn off at around $d\approx 1.6$. 
        }
    \label{fig:shutoff_intensity}
\end{figure}

	In the following we consider an example for a laser for which the overall spatial profile of the
applied pump, $D_0(\mathbf{x}, d)$ evolves non uniformly as a function of the pump parameter $d$. 
As recently pointed out in \cite{liertzer_pump-induced_2012} such a spatially varying pump 
function can strongly influence the laser output in a counterintuitive way, \REFEREE{an effect which has meanwhile also been verified experimentally \cite{brandstetter_reversing_2014}}.
The system we consider to realize such a behavior consists of two
coupled one-dimensional ridge cavities (see inset Fig.~\ref{fig:shutoff_intensity}) which feature strong
loss in the absence of pump. The pump function is defined as follows:
For values of the pump parameter $d$ between $0$ and $1$, only the left
cavity of the system is pumped uniformly with an amplitude that is linearly increasing from zero
(at $d=0$) to a value where the laser is close above threshold (at $d=1$). 
For $d$ between $1$ and $2$ the pump in the left cavity is kept constant (at
the value for $d=1$), while the pump
in the right cavity is linearly increased from zero (at $d=1$) to the same pump
strength as in the left cavity (at $d=2$). Since the overall pump
strength in the cavity steadily increases, one would expect that also
the overall intensity of the laser should increase along this ``pump trajectory'' 
from $d=0$ to $d=2$. Instead, the laser displays an intermittent
shut down within a whole interval of $d$ around $d\approx 1.6$, as shown in 
Fig.~\ref{fig:shutoff_intensity}. 
 
\begin{figure}[tb]
  \includegraphics{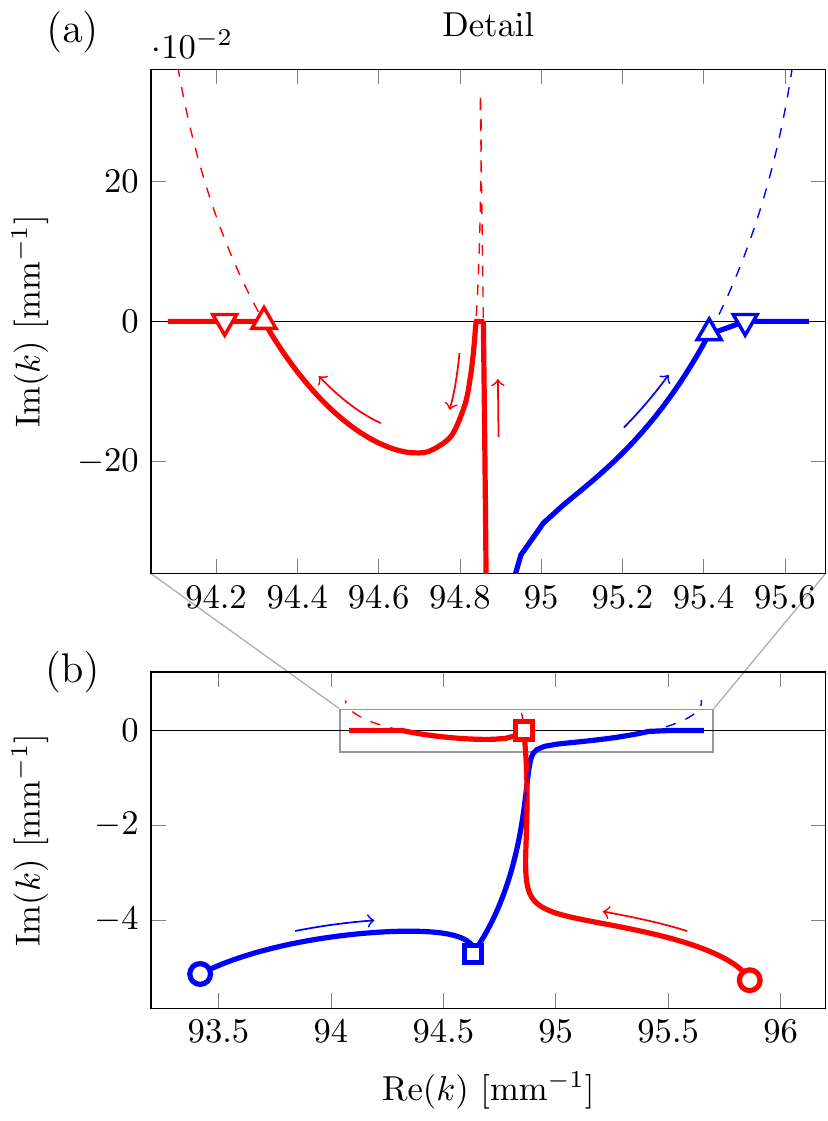}
  \caption{(Color online) Movement of the complex SALT eigenvalues (resonance poles) along the pump trajectory realized in  
  Fig.~\ref{fig:shutoff_intensity}. Solid lines represent the eigenvalues computed with our new solution method and dashed lines represent the   
  solutions in the absence of the non-linear spatial hole burning. Colors (red/blue) are chosen in correspondence with 
  Fig.~\ref{fig:shutoff_intensity}. Our results show that the laser shut down can be associated with an avoided level crossing of the SALT 
  eigenvalues in the complex plane. Details are shown in the top panel (a), where upward triangles mark the eigenvalues where the first mode 
  starts lasing a second time in the course of the pump trajectory. Downward triangles mark the eigenvalues where the second mode starts lasing for
  the first time. In the main panel (b) circles label the eigenvalues at the starting point of the pump trajectory ($d=0$) and squares 
  label the positions of the eigenvalues at the first threshold.}
  \label{fig:shutoff_poles}
\end{figure}

In Ref.~[\citenum{liertzer_pump-induced_2012}] this
shutdown \REFEREE{as obtained with SALT has been quantitatively verified against a traditional FDTD method to show that the solutions are stable and not an artifact of SALT. Furthermore, the shutdown was} attributed to the occurrence of an exceptional point, corresponding to a non-Hermitian 
degeneracy in the TCF eigenvalues $\eta_n$ [see Eq.~(\ref{eq:CF})] when
parametrized over both the outside frequency $k$ and the pump
parameter $d$. In the direct solver, there no longer exists such a two-dimensional parameter space since the frequency $k$ can no longer be
freely adjusted outside the cavity. Instead, the frequency $k$ is already obtained simultaneously with the
corresponding lasing mode.  We can thus expect that the poles associated with the (non)lasing modes reflect, in some form,
their vicinity to the exceptional point through a nontrivial behavior along this pump trajectory.
Indeed, our calculations show that the intermittent laser shut down is realized in terms of an avoided 
crossing between a lasing pole and a non-lasing pole in the complex plane (see 
Fig.~\ref{fig:shutoff_poles}).  Here, the solid lines represent the
solutions of the full SALT while the dashed lines show the movement of
the complex eigenvalues when spatial hole burning is neglected.

In fact, we observe two avoided crossings in this plot. The first one
occurs in the range between $d=0$ (marked as circles in
Fig.~\ref{fig:shutoff_poles}) and $d=1$ (marked as squares). In this case the poles associated with the 
blue and the red mode first attract each other and then undergo an
avoided crossing which pushes the red mode towards and, ultimately, beyond the real axis, i.e., the
 lasing threshold. 

The second avoided crossing occurs in the interval between $d=1$ and $d=2$, where
we observe that the blue pole moves towards the real axis and interacts with the red pole such as to pull it
below the real axis, corresponding to switching this mode off. In a corresponding experiment \cite{brandstetter_reversing_2014}, only the second pole 
interaction can be directly observed in terms of an intermittent laser shut down, followed by a re-emergence of the laser
modes at slightly detuned lasing frequencies.

Figure~\ref{fig:shutoff_poles} also illustrates a crucial point touched on earlier: 
If one neglects the non-linear spatial hole burning interaction (dashed lines) one obtains 
poles in the upper half of the complex plane which violate the causality principle for the dielectric response.
Including spatial hole burning (solid lines), keeps all poles below or on the real axis, as required [see
Fig.~\ref{fig:shutoff_poles}(a)]. Note, that one also observes how the hole-burning interaction influences
the movement of the non-lasing modes in terms of a delayed turn on of the blue mode [see the line between 
the two triangles in Fig.~\ref{fig:shutoff_poles}(a)].

\begin{figure}[h]
  \includegraphics[width=\columnwidth]{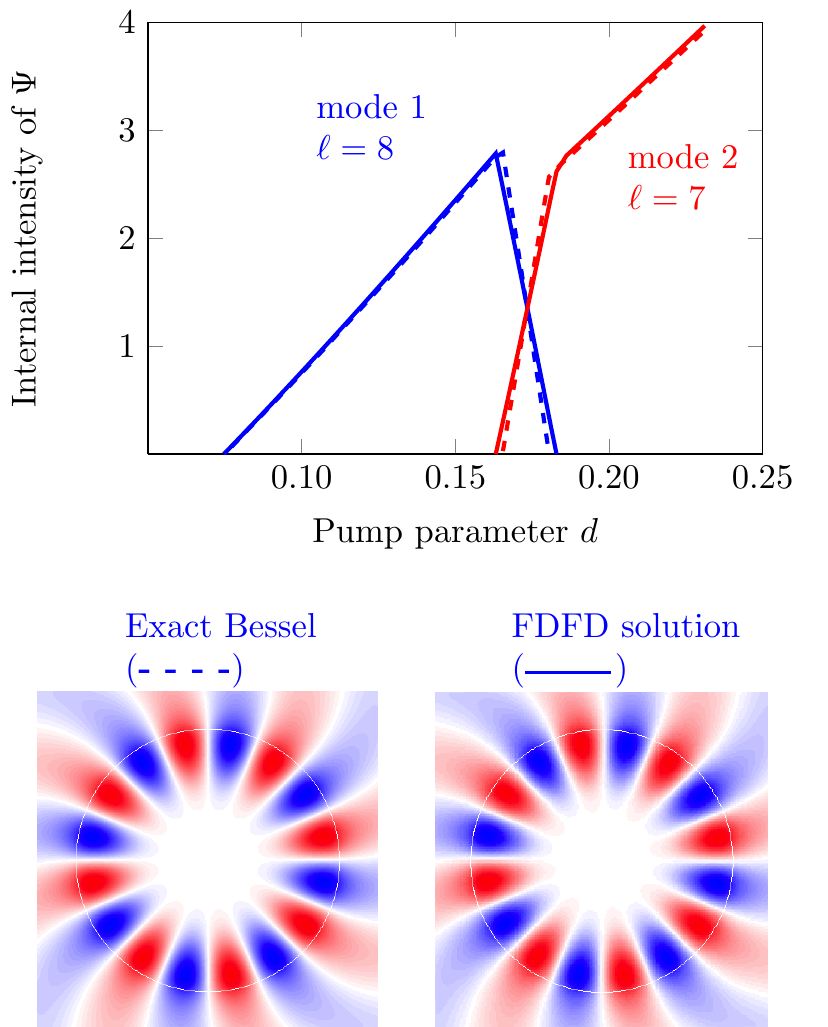}
  \caption{(Color online) Validation of the 2D Newton solver based on FDFD against the CF-state approach (using 20 CF basis states) in a circular 
  cavity with radius $R=100\ \mu\mathrm{m}$ and dielectric index $n=\sqrt{\varepsilon_c}= 2 + 0.01i$. TM-polarized modes are considered and the following gain parameters are used: 
  $\gamma_{\perp} = 10\ \mathrm{mm}^{-1}$, $k_a = 48.3\ \mathrm{mm}^{-1}$.
  Increasing the strength of the uniform pump 
  $D_0(\mathbf{x}, d) = d$ , we encounter strong non-linear modal competition between the first two lasing modes with the result that 
  for sufficiently large pump strength the second lasing mode is found to suppress the first one (see top panel). 
  The internal intensity is defined as the integral over the cavity $ \int | \BPsi(\mathbf{x}) |^2d\mathbf{x}$. 
  The real part of the lasing mode profile $\Psi(\mathbf{x})$ at the first threshold 
  is shown for both the exact Bessel solution ($\Psi \sim e^{-i\ell \theta}$) and for the finite difference solution   
  (see bottom panel, where blue/white/red color corresponds to negative/zero/positive values). As the pump strength is increased, this profile 
  does not change appreciably apart from its overall amplitude. 
}
  \label{fig:cylinder}
\end{figure}
	
	\subsection{Scalability to full-vector 2D and 3D calculations}
	\label{sec:Examples23d}
	In this section we briefly explore the applicability of our solution strategy to 2D and 3D setups by considering 
the following prototypical examples: In the 2D case we investigate a circular dielectric resonator and in the 3D case a photonic-crystal slab.

\begin{figure}[t]
  \includegraphics[width=\columnwidth]{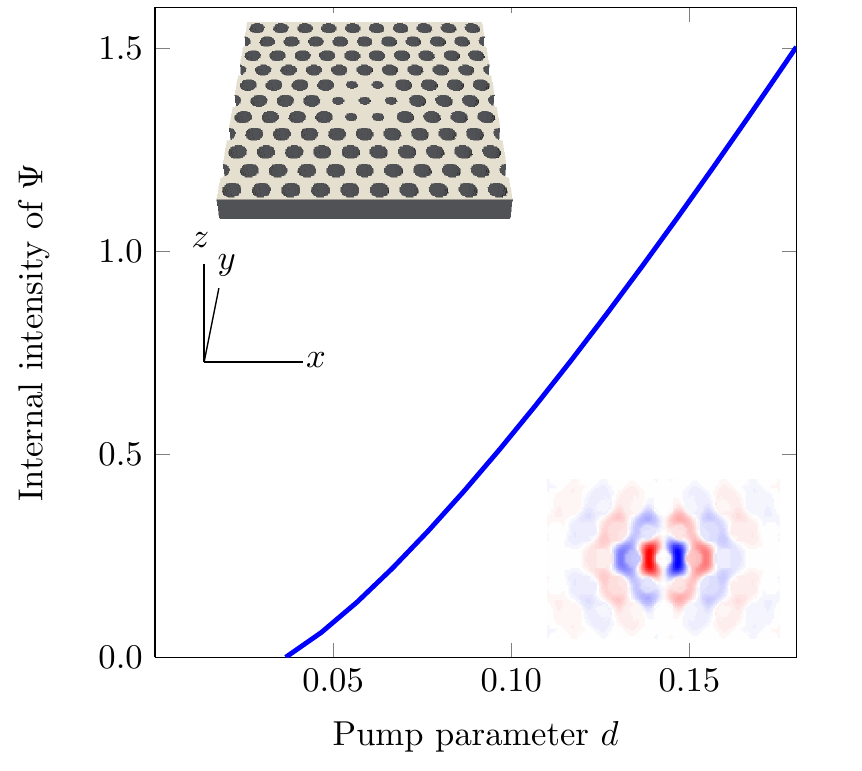}
  \caption{(Color online) 3D calculation of a lasing mode created
by a ``defect'' in a photonic-crystal slab \cite{phc_book}: a period-$a$ hexagonal lattice of air
holes (with $a= 1\ \mathrm{mm}$ and radius $0.3\ \mathrm{mm}$) in a dielectric medium with index $n=\sqrt{\varepsilon_c}=3.4$ with a cavity formed by seven holes of radius
$0.2\ \mathrm{mm}$ in which a doubly-degenerate mode is confined by a photonic bandgap (one of these degenerate modes is
selected by symmetry, see text). The gain has $\gamma_\perp = 2.0\ \mathrm{mm}^{-1}$, 
${k_a=1.5\ \mathrm{mm}^{-1}}$, and non-uniform pump $D_0(\mathbf{x}, d) = f(\mathbf{x}) d$, where the pump profile
 $f(\mathbf{x}) = 1$ in the hexagonal region of height $2\ \mathrm{mm}$ in the $y$-direction,
 and $f(\mathbf{x}) = 0$
outside that region and in all air holes.   
%Even and odd $x=0$ and
%$y=0$ mirror symmetries were imposed to select one of the degenerate modes, with an even $z=0$ mirror
%plane in 3D and the corresponding TE polarization in 2D \cite{phc_book}.
The slab has a finite thickness
$0.5\ \mathrm{mm}$ with air above and below into which the mode can radiate (terminated by PML absorbers). \REFEREE{The inset shows 
magnetic field $H_z$ 
($\sim \partial_x E_y - \partial_y E_x$) of the TE-like mode at the $z=0$ plane.}
    }
  \label{fig:phc}
\end{figure}

In the former situation we study a circular disk with uniform index, which is routinely used in the experiment due to its long-lived resonances
associated with ``whispering gallery modes'' \cite{vahala_optical_2003}. For this system we study lasing based on TM polarized modes and compare 
the Newton method presented here (based on FDFD) with the previously developed CF-state method \cite{tureci_self-consistent_2006,ge_steady-state_2010-1, ge_steady-state_2010}.
Due to the azimuthal symmetry, the resonant TM modes \cite{tureci_self-consistent_2006,internal_disks} are exact solutions of the Bessel equation characterized by an azimuthal phase $e^{\pm i \ell \theta}$ (with 
$\ell$ being an integer angular-momentum quantum number) and subject to outgoing boundary conditions.
Due to the circular symmetry, each of the modes with a given value of $\ell$ comes with a degenerate partner mode, 
characterized by the quantum number $-\ell$.
In the presence of the lasing nonlinearities, a preferred superposition will typically be selected as the
stable solution, e.g.,~the circulating modes $e^{\pm i \ell  \theta}$, rather than the $\sin(\ell  \theta)$ and $\cos(\ell  \theta)$ standing waves. 
The determination
of this stable solution in a degenerate lasing cavity is a complex problem that we plan to address in
future work. For validation and demonstration purposes in this paper, we simply select {\it a priori}
a single solution from each degenerate pair by imposing corresponding symmetry boundary conditions.
In the case of the circular cavity, we choose the circulating modes with a phase $e^{- i \ell  \theta}$ for
comparison with the CF solutions. We obtain these by solving for both the sine and cosine modes 
(using the appropriate boundary conditions at the $x = 0$ and $y = 0$ symmetry planes) and by combining them 
to construct the exponentially circulating mode.

Under these premises, we find that for uniform pump the first mode turns on at $d \approx 0.075$ and increases linearly in 
intensity, as seen in \fig{cylinder}. The second mode turns on at about
twice the pump strength as the first threshold. As the intensity of the second mode increases, we observe a reduction and ultimately a complete 
suppression of the first mode intensity. This mode competition can be attributed to the following two effects:  
The two modes have a significant spatial overlap, such that they compete for the same gain through non-linear spatial hole burning 
which is fully incorporated in SALT. In addition, as being spectrally 
closer to the peak of the gain curve $\gamma(k)$, the second mode can profit
more strongly from the gain in the disk than the first mode. As a result, the second mode prevails against the first mode in this
non-linear competition. This behavior of interaction-induced mode
switching is general and can be found in other laser configurations
and nonlinear media as well \cite{modeswitching2014}. In \fig{cylinder} we show that this behavior is faithfully reproduced with our approach, not only in terms of the
modal intensities as a function of the applied pump (see top panel), but also in terms of the corresponding lasing modes which mirror those 
obtained with the CF-state technique very accurately (see bottom panel).

The second example we consider is a photonic crystal slab with a ``defect'' (see inset \fig{phc}) engineered 
to efficiently trap a mode \cite{sgj_phc_defect2001}. The photonic crystal is formed in a dielectric slab 
by holes which are arranged in a hexagonal lattice and the defect is created by decreasing the 
radius of seven of the holes in the center. In our study, we focus on a TE-like lasing mode, situated
at the defect (spatially) and in the bandgap of the lattice (spectrally). 
To select one of the degenerate standing-wave solutions, we impose even and odd symmetry at $x=0$ and $y=0$,
respectively, as well as an even symmetry at $z=0$.
Staying close to a potential experimental realization, we choose the pump profile $D_0(\mathbf{x}, d)$ to be uniform inside
the slab material's defect region but zero outside and in the air holes. Increasing the overall amplitude of this pump profile, we
find the lasing behavior shown in \fig{phc} (main panel).  \REFEREE{This calculation
was performed with 16 nodes (using one CPU per node) of the Kraken Cray XT5 at the University of Tennessee. 
With $144 \times 120 \times 40$ pixels (the mirror conditions effectively halve these), the total wall-clock time for the computation, from passive resonance at $d=0$ to
lasing above threshold at $d=0.18$, was 5.9 min. Pump steps of $\delta d = 0.02$ were taken, with three to four Newton iterations per pump value.}

\section{Concluding Remarks}

In this paper, we have presented an algorithm for solving the SALT equations which describe the steady-state lasing modes and frequencies of lasers with a free spectral range and a dephasing rate that are both large as compared to the population decay rate and the relaxation oscillation frequency. These conditions are typically satisfied by microlasers with a linear dimension that does not exceed a few hundred wavelengths. Our solution strategy proceeds by a direct discretization using standard methods as FEM or FDFD, without the need for an intermediate CF basis. The resulting increase in efficiency lets our approach scale to complex 2D and 3D lasing structures, which  paves the way for future work in a number of directions. 

First, it is now possible to study lasing in much more complex geometries than could previously be readily simulated, offering the possibility of discovering geometries that induce unexpected new lasing phenomena.  Going one step further, future computations could search a huge space of lasing structures via large-scale optimization (``inverse design''), which has already been applied to the design of \emph{linear} microcavities \cite{frei, santosa, lu_vuckovic}. Since our approach is only more expensive than the solution of linear cavity modes by a small constant factor (e.g., the number of modes and the number of Newton iterations) it will be the ideal tool for this purpose. More complicated gain profiles, lineshapes, and other material properties can easily be incorporated into our approach as well. SALT can, e.g., be coupled to a diffusion equation in order to model the migration of excited atoms in molecular-gas lasers \cite{chua_lasers,cerjan_steady-state_2014}. Based on the mathematical relation of the multimode lasing equations to incoherent vector solitons (Appendix \ref{app:vector_solitons}), we believe that numerical methods commonly used in soliton theory can also be adopted to efficiently solve the multimode SALT equations. Another intriguing direction of research is the development of a more systematic approach to modeling lasers with degenerate linear modes, which requires a technique to evaluate the stability of the solution and evolve an unstable mode to a stable mode.  Finally, many refinements are possible to the numerical methods, such as efficient iterative solvers and preconditioners for the Newton iterations of the lasing modes or criteria to alternate between systematic contour-integral evaluation and simpler Newton-inverse tracking of the non-lasing modes. 
In this sense our approach has more in common with standard sparse discretization methods used to solve other nonlinear PDEs than the CF-basis approach (which is specialized to the SALT problem) and thus opens the door for more outside researchers and numerical specialists to study lasing problems.

\begin{acknowledgments}
  The authors would like to thank the following colleagues for very
  fruitful discussions: S.~Burkhardt, D.~Krimer, X.~Liang, Z.~Musslimani, L.~Nannen, H.~Reid, and J.~Sch\"oberl. S.E., M.L., K.G.M., J.M.M., and S.R.~acknowledge financial support by the Vienna Science and Technology Fund (WWTF) through Project No.~MA09-030, by the Austrian Science Fund (FWF) through Projects No.~SFB IR-ON F25-14 and No.~SFB NextLite F49-P10, by the People Programme (Marie Curie Actions) of the European Union's Seventh Framework Programme (FP7/2007-2013) under REA Grant Agreement No.~PIOF-GA-2011-303228 (Project NOLACOME) as well as computational resources on the Vienna Scientific Cluster (VSC). D.L.~and S.G.J.~were supported in part by the
AFOSR MURI for Complex and Robust On-chip
Nanophotonics (Dr.~Gernot Pomrenke), Grant No.~FA9550-09-1-0704 and by the Army Research Office through the 
Institute for Soldier Nanotechnologies (ISN), Grant No.~W911NF-07-D-0004. A.D.S.~and A.C.~acknowledge the support by the NSF Grants No.~DMR-0908437 and No.~DMR-1307632 and the Yale HPC for their computing resources.
\end{acknowledgments}

\begin{appendix}
 \section{Details for lasing mode solution method}
 \label{app:jacobian}
  In this appendix, we provide further details on setting up the Newton-Raphson iteration for $M$ lasing modes. First, we describe how to fix the phase and normalization for each mode (as mentioned in Sec.~\ref{sec:active_modes}). We choose a point $\mathbf{x}_0$ and a constant unit vector $\left| \mathbf{a} \right| = 1$ such that $\mathbf{a} \cdot \BPsi_\mu (\mathbf{x}_0)$ is nonzero for all lasing modes. This condition is usually satisfied provided that $\mathbf{x}_0$ is neither far outside the cavity nor a point of high symmetry. We then define the quantity $s_\mu \equiv \left| \mathbf{a} \cdot \BPsi_\mu (\mathbf{x}_0)\right|$ and rescale the field such that the \emph{physical} field becomes $s_\mu^{-1} \BPsi_\mu (\mathbf{x})$
and the rescaled field satisfies
\begin{equation}
\label{eq:rescaled}
\mathbf{a} \cdot \BPsi_\mu (\mathbf{x_0}) = 1.
\end{equation}
With this redefinition, the rescaled field $\BPsi_\mu (\mathbf{x})$ has a fixed phase and a normalization that distinguishes it from the trivial solution $\BPsi_\mu (\mathbf{x}) = 0$. Further we treat the quantity $s_\mu$ as a separate unknown that contains the mode's amplitude. The spatial hole-burning [Eq.~(\ref{eq:NLcoupling})] then becomes
\[
D(\mathbf{x},d,\left\{ k_{\nu},s_{\nu},\BPsi_{\nu}(\mathbf{x})\right\} )=\frac{D_{0}\left(\mathbf{\mathbf{x}},d\right)}{1+{\sum \left| \gamma(k_\nu) \BPsi_{\nu}(\mathbf{x})\right|^{2}s_{\nu}^{-2}}}.
\]

Now, we describe how to construct the vector of unknowns $\mathbf{v}$ which, after rescaling, should contain $\BPsi_\mu(\mathbf{x})$, $k_\mu$, and $s_\mu$. First, the discretized fields $\BPsi_\mu(\mathbf{x})$ are described by $N$-component complex vectors $\mathbf{b}_\mu$. The $2N+2$ real unknowns for each mode can then be written in block form as
\begin{equation}
\label{eq:v}
\mathbf{v}^{\nu}=\left(\begin{array}{c}
\mathbf{v}_{1}^{\nu}\\
\mathbf{v}_{2}^{\nu}\\
\mathbf{v}_{3}^{\nu}\\
\mathbf{v}_{4}^{\nu}
\end{array}\right)=\left(\begin{array}{c}
\Re[\mathbf{b}_{\nu}]\\
\Im[\mathbf{b}_{\nu}]\\
k_{\nu}\\
s_{\nu}
\end{array}\right).
\end{equation}
The vector $\mathbf{v}$ we use for the Newton-Raphson method contains all $\mathbf{v}^\mu$ in sequence, since the lasing modes are all coupled together through the spatial hole-burning interaction and thus must be solved  simultaneously. 

Next, we construct the equation vector $\mathbf{f}$ by discretizing the operator $-\nabla \times \nabla \times + k^2_\mu \left[ \varepsilon_c(\mathbf{x}) + \gamma(k_\mu) D \right]$ into a sparse complex matrix $\mathbf{S}_\mu$. In the discrete basis, Eq.~(\ref{eq:SALTsystem}) becomes $\mathbf{S}_\mu  \mathbf{b}_\mu = 0$ which gives $N$ complex scalar equations, and the normalization condition that fixes the phase [Eq.~(\ref{eq:rescaled})] becomes the complex scalar equation $\mathbf{e}^T \mathbf{b}_\mu = 0$, where $\mathbf{e}^T$ is the discrete-basis representation of the vector function consisting of the unit vector $\mathbf{a}$ at point $\mathbf{x}_0$ and zero everywhere else. The real and imaginary parts of these $N+1$ complex equations can be written in block form as
\begin{equation}
\label{eq:f}
\mathbf{f}^{\mu} = \left(\begin{array}{c}
\mathbf{f}_1^{\mu}\\
\mathbf{f}_2^{\mu}\\
\mathbf{f}_3^{\mu}\\
\mathbf{f}_4^{\mu}
\end{array}\right)
= \left(\begin{array}{c}
\Re\left[\mathbf{S}_{\mu}\mathbf{b}_{\mu}\right]\\
\Im\left[\mathbf{S}_{\mu}\mathbf{b}_{\mu}\right]\\
\Re\left[\mathbf{e}^{T}\mathbf{b}_{\mu}\right]-1\\
\Im\left[\mathbf{e}^{T}\mathbf{b}_{\mu}\right]
\end{array}\right).
\end{equation}
The vector $\mathbf{f}$ we use for the Newton-Raphson method contains all $\mathbf{f}^\mu$ in sequence, due to the intermodal coupling. 

Finally, we describe how to construct the (real) Jacobian matrix $\mathcal{J}$ (which is real), which consists of $M^2$ blocks $\mathcal{J}^{\mu \nu}$ that each have size $2N+2$ and have the block form
\[
\mathcal{J}_{ij}^{\mu \nu} = \frac{\partial \mathbf{f}_i^\mu}{\partial \mathbf{v}_j^\nu}.
\]
We explicitly construct these blocks by taking derivatives of column blocks of $\mathbf{f}_i^\mu$ with respect to row blocks of $(\mathbf{v}_j^\nu)^T$, as defined in Eqs.~(\ref{eq:v}) and (\ref{eq:f}). 
First, we see that $\mathcal{J}_{31}^{\mu \nu} = \mathcal{J}_{42}^{\mu \nu} = \mathbf{e}^T$, while all other blocks of $\mathcal{J}_{ij}^{\mu \nu}$ with $i=3,4$ are zero. Second, we have the columns
\[
\left(\begin{array}{c}
\mathcal{J}_{13}^{\mu\nu}\\
\mathcal{J}_{23}^{\mu\nu}
\end{array}\right)=\left(\begin{array}{c}
\Re\\
\Im
\end{array}\right)\left[\frac{\partial\mathbf{S}_{\mu}}{\partial k_{\nu}}\mathbf{b}_{\mu}\right]
\]
and
\[
\left(\begin{array}{c}
\mathcal{J}_{14}^{\mu\nu}\\
\mathcal{J}_{24}^{\mu\nu}
\end{array}\right)=\left(\begin{array}{c}
\Re\\
\Im
\end{array}\right)\left[\frac{\partial\mathbf{S}_{\mu}}{\partial s_{\nu}}\mathbf{b}_{\mu}\right],
\]
where the derivatives of $\mathbf{S}_\mu$ are diagonal complex matrices that can be obtained straightforwardly by discretizing the same derivatives of the complex scalar function $k^2_\mu \left[ \varepsilon_c(\mathbf{x}) + \gamma(k_\mu) D \right]$. (In the case that exact outgoing radiation conditions are used for $\mathbf{S}_\mu$, the matrix for $-\nabla \times \nabla \times$ may also depend on $k_\mu$ and this dependence must also be included in the derivative.)

Finally, the remaining blocks are given by
\[
\left(\begin{array}{cc}
\mathcal{J}_{11}^{\mu\nu} & \mathcal{J}_{12}^{\mu\nu}\\
\mathcal{J}_{21}^{\mu\nu} & \mathcal{J}_{22}^{\mu\nu}
\end{array}\right)=\left(\begin{array}{cc}
\Re & -\Im\\
\Im & \Re
\end{array}\right)\mathbf{S}_{\mu}\delta_{\mu\nu}+\mathbb{S}_{\mu\nu}
\]
where $\delta_{\mu \nu}$ is the Kronecker $\delta$, and $\mathbb{S}_{\mu \nu}$ is the matrix discretization of the real $6 \times 6$ tensor function
\[
2k_\mu^2 \gamma (k_\mu) \frac{\partial D}{\partial \left| \BPsi_\nu(\mathbf{x})\right|^2} 
\BPsi_\mu(\mathbf{x}) \otimes \BPsi_\nu(\mathbf{x})
\]
with the outer product $\otimes$ taken over the real and imaginary parts of the vector components of $\BPsi(\mathbf{x})$.

 \section{FEM formalism}
 \label{app:FEM}
  In this appendix we provide details on how to implement our SALT solution strategy with a high order finite element method
($hp$-FEM) 
  \cite{demkowicz_computing-2007, solin_partial_2005}. Most importantly, our approach
does not depend on this specific discretization method, but it entails
several significant advantages. Specifically, $hp$-FEM can handle highly complex, irregularly shaped geometries 
and the higher order discretizations tend to exponentially increase
the accuracy of the computations (if all of the boundary discontinuities and corner singularities are properly taken into account). 

In order to obtain the discretized formulation, we truncate the open problem to a bounded computational domain $\Omega$. 
Then we multiply
Eq.~(\ref{eq:SALTsystem}) with an arbitrary test function $v$ and
integrate over the domain of the cavity. 
Using the Green's formula
this leads to the weak formulation
\begin{equation}
   \label{eq:SALTweak}
   \begin{aligned}
   \int_\Omega (\nabla \times \BPsi_\mu)\cdot (  \nabla \times v) + k_\mu^2 \int_\Omega 	
   \varepsilon_\mu({\bf x}, \{k_\nu, \BPsi_\nu\}) \BPsi_\mu\cdot  v & \\
   - \int_{\partial \Omega} [\vec{n}  \times (\nabla \times \BPsi_\mu)] \cdot v &= 0,
   \end{aligned} 
\end{equation}
where $\vec{n}$ denotes the outer normal vector at the boundary~$\partial\Omega$.
The boundary term $\vec{n}  \times (\nabla \times \BPsi_\mu)$ has to be replaced by a term incorporating the radiation condition at infinity. Formally, this can be done by $\vec{n}  \times (\nabla \times \BPsi_\mu) = DtN(\BPsi_\mu)$, where $DtN$ is the Dirichtlet-to-Neumann operator \cite{nedelec01}.
In one dimension and in two dimensions for TE modes the Maxwell equations reduce to the well-known scalar Helmholtz equation. 
Furthermore, in the one-dimensional case the boundary integral can be simply replaced by \begin{equation}
\label{eq:boundary_integral}
\int_{\partial \Omega} \partial_{\vec{n}}\BPsi_\mu v = -ik \int_{\partial \Omega} \BPsi_\mu v.
\end{equation}
In higher dimensions an appropriate representation of the open boundary becomes more sophisticated, as mentioned in Sec.~\ref{sec:outgoing}. For a detailed discussion see \cite{monk2003finite}, but Eq.~\eqref{eq:boundary_integral} can also be simply used as the first-order approximation of the $DtN$ operator.

%Then 
The unknown laser modes $\BPsi_\mu$ are sought as a linear
combination of element basis functions $\{\varphi_j\}$ such that
\begin{equation}
   \label{eq:ansatz}
   \BPsi_\mu(\mathbf{x}) = \sum_{j=1}^{N} b^\mu_j \varphi_j(\mathbf{x})
\end{equation}
where $\{ \varphi_j\}_{j=1}^{N}$ are piecewise polynomials with local
support and $N$ is the number of degrees of freedom of the system. For
more details on the choice of such a basis, based on high-order
elements, we refer to 
 \cite{demkowicz_computing-2007}.
We use the notation $X = \{k_\nu, { \bf b_\nu}\}_{\nu=1}^{M}$ with a
complex FEM-coefficient vector ${ \bf b_\nu}\defined \left(b^\nu_1,\ldots,
  b^\nu_{N}\right)$. Inserting the ansatz Eq.~(\ref{eq:ansatz}) into
Eq.~(\ref{eq:SALTweak}), extracting sums out of the integrals, and
assembling the contributions for all elements we arrive at the
following finite-element scheme in matrix form:
\begin{equation}
   \label{eq:SALT_FEM}
   \big[-{\bf L} + ik_\mu {\bf R} + k_\mu^2 {\bf M}^{\varepsilon_c} + k_\mu^2
   \gamma(k_\mu){\bf Q}(X)\big]{\bf b_\mu}  = 0,
\end{equation} 
where the sparse $N \times N$-matrices ${\bf L}, {\bf R} ,{\bf
  M}^{\varepsilon_c}, {\bf Q}(X)$ are the stiffness matrix
$$ {\bf L} \defined \left\{\begin{array}{cc}
\displaystyle \left(\int_{\Omega} \nabla \varphi_i \cdot \nabla \varphi_j dx\right)_{i,j}, & \text{ for } d=1,2,\\
\displaystyle  \left(\int_{\Omega} (\nabla \times \varphi_i) \cdot (\nabla \times  \varphi_j) dx\right)_{i,j}, & \text{ for } d=3,\\
\end{array}\right.$$
corresponding to the Laplacian or curl-curl term, respectively, the mass matrix
$${\bf M}^{\varepsilon_c}\defined \left(\int_{\Omega}  \varepsilon_c(x) \varphi_i\cdot \varphi_j dx\right)_{i,j} $$
containing the passive dielectric function, the matrix 
$${\bf R} \defined \left(\int_{\partial \Omega} \varphi_i \cdot\varphi_j d\sigma_{\mathbf{x}} \right)_{i,j},$$
which only involves the boundary elements and incorporates the
outgoing boundary condition \eqref{eq:boundary_integral}, and the nonlinear contribution
$${\bf Q}(X) \defined \left( \int_{\Omega}
  \REFEREE{ \frac{D_0(x,d)\varphi_i(x) \varphi_j(x)}{1+\sum_\nu
    |\gamma(k_\nu)\sum_l b^\nu_l \varphi_l(x)|^2} } dx\right)_{i,j}, $$
which accounts for the nonlinear coupling including the spatial hole burning effect.

 \section{FDFD formalism}
 \label{app:FDFD}
   For finite-difference calculations shown in the main text, the discretization code implemented in \cite{xd_optexp, xd_thesis} was used. The complex electric fields
$\BPsi_\mu$ were discretized on an $N_x \times N_y \times N_z$ pixel grid of equally spaced points with
the $- \nabla \times \nabla \times$ operator being conveniently discretized using second-order centered differences on a Yee lattice \cite{taflove}.
To impose outgoing boundary conditions, additional pixels of PML were
added at the boundaries with the appropriate absorption, as explained in Sec.~\ref{sec:outgoing}.
For each mirror symmetry in a geometry, we were able to halve the computational domain by replacing the PML at the lower
walls with the corresponding boundary conditions
of the mirror plane. Furthermore, for the cases of TM ($E_{x,y} = 0$) and TE polarization ($E_z = 0$), 
the problem size can be reduced by factors of $3$ and $3/2$, respectively, by 
projecting $\mathbf{S}_\mu$ and $\mathbf{b}_\mu$ into the nonzero field components only. 
Additionally, 2D calculations were performed by setting $N_z=1$ and the boundary condition in the $z$ direction to be
periodic. 1D calculations were performed by doing so for both the $z$ and $y$ directions.

 \section{Contour integral method}
 \label{app:beyn}
 This section reviews the algorithm for solving a nonlinear EVP of the form ${\bf Sb}=0$ as discussed in  Ref.~[\citenum{beyn_integral_2012}]. 
The corresponding inverse matrix ${\bf S}^{-1}$  can generically be written as
$$ {\bf S}^{-1}(k) = \sum_n \frac{1}{k - {k}_n} {\bf v}_n {\bf w}^H_n + {\bf H}(k),$$
where ${k}_n$ are the desired complex eigenvalues and ${\bf v}_n,
{\bf w}_n$ are the corresponding left and right eigenvectors. The residual term ${\bf H}$ is holomorphic. Using the
residue theorem the following two matrices can be defined:
$$ {\bf A}_0 \defined \displaystyle \frac{1}{2\pi i}\oint_\mathcal{C}  {\bf
  S}^{-1}(k) dk =  \sum_n   {\bf v}_n {\bf w}^H_n  = {\bf VW}^H,$$ 
$$ {\bf A}_1 \defined \frac{1}{2\pi i}\oint_\mathcal{C}  k{\bf S}^{-1}(k) dk =
\sum_n {k}_n {\bf v}_n {\bf w}^H_n = {\bf V K W}^H,$$ where $\bf K$ is
a diagonal matrix with the diagonal entries corresponding to all the
poles of the inverse matrix ${\bf S}^{-1}$ inside the contour
$\mathcal{C}$ which in turn are the eigenvalues of ${\bf S}$.
Before we show how the desired matrix ${\bf K}$ is computed from the
two matrices ${\bf A}_0$ and ${\bf A}_1$, we discuss the realization of
the contour integration which is obtained by numerical
quadrature. Very fast (i.e., exponential) convergence is achieved with
the trapezoidal rule
\cite[Theorem 9.28]{kress_numericalanalysis_1998}, if the contour is an
analytic curve such as a circle or an ellipse.
Moreover,  
inverting the large matrix ${\bf S}$ for each quadrature point is
numerically expensive and may even be infeasible given that the
inverses are fully populated.
This can be remedied by an approximation scheme that exploits the fact
that the rank of the matrices ${\bf A}_0$ and ${\bf A}_1$ is given by
the number of eigenvalues inside the contour and is thus very small
compared to $N$: For a random matrix ${\bf M}\in \mathbb{R}^{ N \times
  l}$, we merely evaluate ${\bf S}^{-1} {\bf M}$ at each quadrature
point on the contour, i.e., we reduce the computational cost to the
solution of $l$ linear systems for each quadrature point.  The
parameter $l$ has to be selected slightly larger than the expected
size of the number of eigenvalues inside the contour.
To obtain the matrix ${\bf K}$, we first compute the (reduced) singular value decomposition (SVD) of 
$${\bf A}_0{\bf M} = {\bf V}_0 {\bf \Sigma}_0 {\bf W}_0.$$ 
Then we define  the
matrix $ {\bf B}\defined {\bf V}_0{\bf A}_1{\bf W}_0{\bf \Sigma}^{-1}_0$
and observe that ${\bf K}$ and $ {\bf B}$ are similar, i.e., ${\bf K} = {\bf P}{\bf B}{\bf P}^{-1}$ for some matrix ${\bf P}$. 
Therefore their eigenvalues are the same such that the desired
eigenvalues ${k}_n$ can be obtained from the reduced eigenvalue
problem $$ {\bf Bx}_n = {k}_n {\bf x}_n.$$
In this short sketch we have assumed that $l$ is exactly
the number of eigenvalues inside the contour so that $\Sigma_0^{-1}$ exists; 
if $l$ is larger, the SVD of ${\bf A_0} {\bf M}$ has to be replaced with 
a rank-revealing variant. In total, the algorithm involves 
a (dense) $l \times l$ eigenvalue problem, an SVD of an $N \times l$ matrix, 
and $q \times l $ sparse linear solves, where $q$ is the number of quadrature points 
for the contour integration. 
Consequently, the bottleneck of a large scale eigenvalue problem is
essentially shifted to the solution of perfectly parallelizable linear
systems.
The contour integration method assumes that the contour does not pass
through eigenvalues; since eigenvalues close to the contour could
affect the convergence of the quadrature scheme, we compute the
residuum of the computed eigenvalues found by the algorithm before
proceeding.
 
Our computations are performed with ellipsoidal contours and the 
trapezoidal rule. If contours are desired that are no longer analytic but
only piecewise analytic, then the trapezoidal rule should be replaced by 
other exponentially convergent schemes such as Gaussian or Clenshaw Curtis 
quadrature.   
However, with quadrature error control in place, this method guarantees to find all eigenvalues inside the contour, but avoids spurious eigenvalues at $k_a+i\gamma_\bot$. Other eigenvalue solvers either compute all eigenvalues which is neither realistic nor necessary 
or rely on local convergence properties.

 \section{Multimode Lasing and Vector Solitons}
  \label{app:vector_solitons}
In this appendix we discuss an interesting mathematical connection between the SALT lasing equations in the multimode regime and the nonlinear incoherent ``vector solitons''  in photorefractive media~\cite{Christodoulides1996}. The noninstantaneous nonlinearity in such media allows more than two components of solitons to be self-trapped and thus to form vector solitons based on the mutual incoherence between their various components (the term ``vector'' refers here to the locked components that propagate together). 
Following~\cite{Christodoulides1996}, the normalized multimode soliton equations for the scalar electric-field envelopes $U_n(x,y,z)$ of $M$ interacting beams are:
 \begin{equation}
 \label{eq:soliton}
 i\frac{\partial U_n}{\partial z} + \frac{1}{2} \nabla^2 U_n - \frac{\beta(1+\rho)U_n}{1+\sum_{m=1}^M |U_m|^2} = 0,
 \end{equation}
where $\rho$  is the total intensity at infinity and $\beta$ is the peak nonlinear index.  Equation~\eqref{eq:soliton} describes $M$ coupled beams in a saturable optical photorefractive medium. Such an interaction can form vector solitons that consist of two or more components mutually self-trapped in the nonlinear medium. In the small-intensity regime (Kerr limit), and when $M=2$, the resulting governing equations describe the so-called Manakov solitons \cite{Manakov1974,Kang1998}, which in 1+1 dimensions are known to be integrable. 
The stationary soliton solutions have 
%Since we are interested for soliton solutions we look for stationary solutions of 
the form $U_n(x,y,z)=V_n(x,y)e^{i\lambda_n z}$ with the soliton eigenvalues  $\lambda_n$ being real numbers. Substituting this ansatz into Eq.~\eqref{eq:soliton} leads to the following nonlinear eigenvalue problem:
 \begin{equation}
 \label{eq:stationary_soliton}
 \nabla^2 V_n -  \frac{2\beta(1+\rho)V_n}{1+\sum_{m=1}^M |V_m|^2} = 2\lambda_nV_n
 \end{equation}
subject to the boundary conditions $\lim_{x,y\to\infty} V(x,y)=0$. 
By comparing with Eq.~\eqref{eq:SALTsystem}, we stress that Eq.~\eqref{eq:stationary_soliton} is an eigenvalue problem defined on the whole real line $x\in (-\infty, +\infty)$, while the laser problem is restricted to the domain of a finite cavity length, $x\in (-L, +L)$. Hence, as a result of the different boundary conditions Eq.~\eqref{eq:stationary_soliton} admits a continuum family of soliton solutions for any nonzero positive value of the soliton eigenvalue $\lambda_\mu$, whereas Eq.~\eqref{eq:SALTsystem}  admits a discrete family of solutions corresponding to different lasing frequencies that are determined self-consistently.
Note also that in extension of the above soliton equations, in SALT not only the eigenvectors but also the eigenvalues appear non-linearly. In spite of such characteristic differences
we can envision applying various numerical methods developed in the field of nonlinear optics and soliton theory to solve efficiently the multimode SALT equations. In particular, the most commonly used techniques used to numerically determine vector soliton solutions are the multi-dimensional Newton-Raphson method \cite{Efremidis2003}, the self-consistent method \cite{Cohen2003} and the recently developed spectral renormalization method \cite{Ablowitz2005}. 

\end{appendix}

% \bibliography{references_sed}{}


\end{document}